\documentclass{aa}  

\usepackage{graphicx}
\usepackage{amsmath}
\usepackage{booktabs}
\usepackage{hyperref}
\usepackage{txfonts}
\usepackage{subcaption}         
\usepackage{lscape}             
\usepackage{placeins}           
                                
\begin{document}

   \title{Low-luminosity Wolf–Rayet stars: a model–data comparison}

%




\author{
Siyu Wu\inst{1,2,3}
\and Zhi Li\inst{1,4,5}\fnmsep\thanks{Corresponding author. Email: \texttt{lizhi@ynao.ac.cn}}
\and Yan Li\inst{1,4,5,6,7}
\and Vladimir Lipunov\inst{8}
\and Guillermo Garc\'{\i}a-Segura\inst{9}
\and Youdong Hu\inst{10}
\and Alberto J. Castro-Tirado\inst{2}
\and Maria Gritsevich\inst{2,11,12}
\and Ignacio P\'erez-Garc\'{\i}a\inst{2}
\and Mar\'{\i}a D. Caballero-Garc\'{\i}a\inst{2}
\and Rub\'en S\'anchez-Ram\'{\i}rez\inst{2}
\and Sergiy Guziy\inst{2}
\and Emilio~J.~Fern\'andez-Garc\'{\i}a\inst{2}
\and Bin-Bin Zhang\inst{13,14}
}

\institute{
Yunnan Observatories, Chinese Academy of Sciences, Kunming 650216, People's Republic of China
\and Instituto de Astrofísica de Andalucía, Consejo Superior de Investigaciones Científicas (IAA-CSIC), Glorieta de la Astronomía, s/n, 18080 Granada, Spain
\and Department of Physics and Mathematics, University of Granada, 18071 Granada, Spain
\and Key Laboratory for Structure and Evolution of Celestial Objects, Chinese Academy of Sciences, People's Republic of China
\and International Centre of Supernovae, Yunnan Key Laboratory, Kunming 650216, P. R. China
\and University of Chinese Academy of Sciences, Shijingshan District, Beijing, 100049, People's Republic of China
\and Center for Astronomical Mega-Science, Chinese Academy of Sciences, Beijing 100012, People's Republic of China
\and Lomonosov Moscow State University, Moscow, universitetskiy prospekt, 13, 119192, Russia 
\and
Instituto de Astronom\'{\i}a, Universidad Nacional Aut\'{o}noma de M\'{e}xico (IA-UNAM), carretera Tijuana-Ensenada, km. 107, C.P. 22860 Ensenada, Baja California, M\'{e}xico
\and
Guangxi Key Laboratory for Relativistic Astrophysics, School of Physical Science and Technology, Guangxi University, Nanning 530004, China
\and
Faculty of Science, University of Helsinki, Gustaf Hallstr\"{o}min katu 2, FI-00014 Helsinki, Finland
\and
Institute of Physics and Technology, Ural Federal University, Mira str. 19, 620002 Ekaterinburg, Russia
\and
School of Physics, Nanjing University, 22 Hankou Road, Nanjing 210023, China
\and
Purple Mountain Observatory, Chinese Academy of Sciences, Nanjing 210023, China
}

\date{Received September 30, 20XX}
 

\abstract
{Galactic Wolf--Rayet (WR) stars reported at comparatively low luminosities provide stringent tests of massive-star evolution, because their luminosities, temperatures, surface compositions, and wind densities depend on mixing and mass-loss history.}
{We investigate which low-luminosity WR stars can be reproduced by rotating single-star models at approximately solar metallicity, and which observables drive the remaining discrepancies. We distinguish mismatches in luminosity from those in temperature or radius, surface composition, and WR-like wind strength.}
{We compare published atmosphere-derived parameters for Galactic low-luminosity WR stars with Modules for Experiments in Stellar Astrophysics (MESA) evolutionary tracks based on the $k$--$\omega$ framework at $Z=0.02$. The comparison uses the Hertzsprung--Russell (HR) diagram position, subtype and surface-composition constraints, and wind-budget diagnostics, including $v_\infty/v_{\rm esc}$, the wind-momentum efficiency $\eta_{\rm wind}$, and the mechanical wind luminosity. We test track families with different red-supergiant (RSG) and WR-phase mass-loss prescriptions.}
{Enhanced RSG mass loss lowers the minimum initial mass for WR-star formation to about $18\,M_\odot$, and the nitrogen-sequence (WN) stars are the least problematic cases. The transitional WN/C (WNC) stars are the most restrictive, because they require a short-lived transitional surface composition together with the observed HR diagram position and wind properties. For the low-luminosity late-type WC (WCL) stars WR\,92 and WR\,119, empirically calibrated WR winds help the tracks reach the faint luminosity range, but the selected model points remain too hot and too compact. The early-type WC (WCE) star WR\,38 is matched more closely in the HR diagram, although it requires the highest wind efficiency.}
{The low-luminosity WR problem is therefore not only a luminosity problem. Revised wind prescriptions alleviate part of the tension, but they do not fully solve the discrepancies in temperature, radius, surface composition, and wind density. Low-luminosity WNC and WCL stars remain key benchmarks for WR mixing, WR-phase mass loss, and possible post-interaction channels.}


   \maketitle


\section{Introduction}
\label{sec:introduction}

Wolf--Rayet (WR) stars are evolved massive stars with dense, fast winds that produce broad emission-line spectra and dominate the emergent radiation field of the stellar surface \citep{Crowther2007}.
In the classical spectroscopic scheme, nitrogen-sequence WR (WN) stars are characterized by strong N and He emission and are commonly interpreted as objects exposing CNO-cycle products, whereas carbon-sequence WR (WC) and oxygen-sequence WR (WO) stars show C and O emission associated with the exposure of He-burning products \citep{Crowther2007,Hamann2006,Sander2012}.
Because the WR phase removes and reveals distinct layers of the star, WR populations provide stringent tests of stellar-evolution physics, in particular the interplay between internal mixing and mass loss \citep{Crowther2007}.

Hydrogen-deficient WR stars have long been considered potential progenitors of stripped-envelope core-collapse supernovae (SNe), especially SNe~Ib/Ic \citep[e.g.,][]{Crowther2007,2022ApJS..262...26L}.
Related channels may also lead to SNe~IIb when a thin H envelope remains, or to SNe~Ibn when the explosion occurs in dense He-rich circumstellar material \citep{2007ApJ...657L.105F,2007Natur.447..829P,2022ApJ...927...25M}.
However, binary interaction can also remove the H-rich envelope and produce hot stripped stars with lower luminosities and weaker winds \citep[e.g.,][]{Eldridge2017,Gotberg2018}.
Low-luminosity WR stars therefore provide useful tests of how mass loss, internal mixing, and possible binary stripping shape the final pre-supernova population.

Empirically, the physical properties of WR stars are most directly constrained by quantitative non-local thermodynamic equilibrium (non-LTE) atmosphere analyses of optical/ultraviolet (UV) spectra.
For Galactic WR stars, recent homogeneous analyses have provided distributions of luminosity, stellar temperature, wind parameters, and (where accessible) surface abundances for both WC and WN populations \citep[e.g.,][]{Sander2012,Hamann2019,Sander2019}.
These data sets enable model--data comparisons that go beyond subtype counts and instead confront the predicted locations of WR stars in the Hertzsprung--Russell (HR) diagram, their inferred evolutionary sequences, and their wind properties.

A long-standing tension concerns the low-luminosity regime of the WR population.
Several Galactic WC and transitional WN/C (WNC) objects have been inferred to have relatively low luminosities compared to expectations from standard single-star evolutionary pathways at (near-)solar metallicity \citep[e.g.,][]{Sander2012,Sander2019}.
At face value, such objects challenge the predicted minimum initial mass for producing WC-like spectra, as well as the predicted duration of the transition from WN to WC.
However, inferred luminosities and mass-loss rates can be biased by distance and extinction uncertainties, wind clumping, unresolved companions, and spectral dilution by binary products \citep[e.g.,][]{Gotberg2017,Shenar2020}.

From the modeling perspective, the faint-WR problem is tightly connected to uncertain physical ingredients in one-dimensional (1D) stellar evolution.
Single-star evolutionary grids that include rotation and calibrated mass-loss prescriptions (e.g., Geneva) predict WR subtype lifetimes and the dependence of the WR population on metallicity \citep[e.g.,][]{Ekstrom2012,Georgy2012}. The inferred final masses of WR stars are also sensitive to the adopted mass-loss prescription; in particular, older models with stronger WR winds could strip even very massive stars with zero-age main-sequence masses, $M_{\rm ZAMS}\simeq 60\,M_\odot$, down to final masses of only a few solar masses \citep{Langer1994}.
In parallel, tracks computed with Modules for Experiments in Stellar Astrophysics (MESA) and related isochrone frameworks provide flexible, widely used baselines for massive-star evolution and population studies \citep[e.g.,][]{Choi2016,Paxton2011,Paxton2013,Paxton2015,Paxton2018,Paxton2019}.
A key uncertainty shared by these approaches is the efficiency of mixing near convective boundaries (often parameterized via diffusive overshooting), which can reshape internal chemical gradients and alter the timing and duration of the WN$\rightarrow$WNC$\rightarrow$WC transition \citep[e.g.,][]{Herwig2000}.

In this paper we focus on low-luminosity Galactic WR stars by directly comparing observationally inferred parameters to predictions from stellar-evolution models.
As the physical backbone of our analysis, we adopt the \citet{Li2023} MESA framework at $Z=0.02$, which implements a turbulence-motivated $k$--$\omega$ treatment of convection and convective-boundary mixing.
In practice, however, our star-by-star comparison is carried out against a small set of related track families within this broader framework, including different wind prescriptions, enhanced-mass-loss variants, and selected higher-rotation branches.
Our goal is to assess whether the observed HR diagram locations, subtype classifications, and wind properties of low-luminosity WR objects remain consistent with single-star evolution under plausible assumptions, and to identify cases where dominant systematics and/or additional evolutionary channels (e.g., binary stripping) are likely required.

This paper is organized as follows.
Section~\ref{sec:obs} describes the observational sample and adopted stellar and wind parameters.
Section~\ref{sec:models} summarizes the evolutionary models and the comparison methodology.
Sections~\ref{sec:results} and \ref{sec:discussion} present the model--data comparison and discuss implications for mixing, mass loss, binary contributions, and a supplementary Gaia-based kinematic context.
We summarize our conclusions in Section~\ref{sec:conclusions}.

\section{Observational sample and stellar parameters}
\label{sec:obs}

\subsection{Sample definition}
\label{subsec:sample}

Our observational sample is drawn from homogeneous non-LTE atmosphere analyses of Galactic WR stars.
We adopt the WN parameters from \citet{Hamann2019}, the WC parameters and the WNC parameter of WR~58 from \citet{Sander2012,Sander2019}, and the parameters of WR~121--16 from \citet{zhang2020}, using the earlier Galactic WN Potsdam Wolf-Rayet (PoWR) analysis of \citet{Hamann2006} as a reference for the model-dependent distinction between H-bearing late-type WN (WNL) and H-free early-type WN (WNE) atmosphere solutions.
Throughout this paper, the observational quantities refer to the values reported in those studies.

The sample contains nine stars satisfying all three selection criteria. This number reflects the scarcity of Galactic WR stars with both low inferred luminosity and homogeneous atmosphere solutions; relaxing either the luminosity threshold or the single-star requirement would introduce systematics that are difficult to control uniformly. The sample nonetheless spans the complete WN–WNC–WC subtype sequence in the low-luminosity regime, making it appropriate for a staged model–data comparison focused on physical diagnostics rather than population statistics.

The main focus of this work is the low-luminosity WNC/WC regime, while the WN stars are retained as a reference sample for comparison across the broader WR parameter space.
Objects flagged in the source papers as potentially affected by unresolved companions are not excluded a priori. Nevertheless, their individual properties and evolutionary interpretations should be treated with caution.

\subsection{Adopted parameters and uncertainties}
\label{subsec:obsparams}

For each star we use the published luminosity $\log(L/L_\odot)$, stellar temperature $T_\ast$, terminal wind speed $v_\infty$, mass-loss rate $\dot{M}$, and subtype-dependent surface-composition diagnostics.
Here $T_\ast$ is the characteristic temperature defined in the atmosphere analyses at the corresponding model radius $R_\ast$.
Mass-loss rates are quoted as $\log\dot{M}$ in $M_\odot\,{\rm yr^{-1}}$, and wind speeds in ${\rm km\,s^{-1}}$.

In the subsequent model comparison, $(L, T_\ast)$ define the observational location of each star in the HR diagram, while subtype and surface-composition information provide an independent check on evolutionary phase.
For uncertainties, we adopt the values quoted in the original atmosphere analyses.
Potential systematics related to distance, extinction, clumping, and unresolved companions are considered only in the interpretation of individual objects, especially in the faintest part of the sample.


\subsection{Escape-speed and wind-scaling diagnostics}
\label{subsec:vesc}

For the representative evolutionary solutions selected later, we evaluate a small set of wind-scaling quantities using the corresponding model values of $M$ and $R$.
This avoids mixing atmosphere-based stellar parameters with masses and radii defined in a non-uniform way across different studies.

We first define the classical escape speed as
\begin{equation}
    v_{\rm esc} = \sqrt{\frac{2GM}{R}} \, .
\end{equation}
When the surface H mass fraction is available, we also consider the effective escape speed corrected for electron-scattering acceleration,
\begin{equation}
    v_{\rm esc,eff} = \sqrt{\frac{2GM(1-\Gamma_{\rm e})}{R}} \, ,
\end{equation}
where
\begin{equation}
    \Gamma_{\rm e} \equiv \frac{\kappa_{\rm e}L}{4\pi cGM}, \qquad
    \kappa_{\rm e} \simeq 0.2(1+X_{\rm H})~{\rm cm^2\,g^{-1}} \, ,
\end{equation}
and $X_{\rm H}$ is the surface H mass fraction \citep{LamersCassinelli1999}.

We then use the ratios $v_\infty/v_{\rm esc}$ and, where applicable, $v_\infty/v_{\rm esc,eff}$, together with the wind-momentum efficiency
\begin{equation}
\eta_{\rm wind} \equiv \frac{\dot{M}v_\infty}{L/c} \, ,
\end{equation}
and the wind mechanical luminosity
\begin{equation}
L_{\rm wind} = \tfrac{1}{2}\dot{M}v_\infty^2 \, .
\end{equation}
These quantities are used later as comparative diagnostics of the momentum and energy requirements of the wind, rather than as direct fitting constraints \citep{LamersCassinelli1999,Crowther2007,Puls2008}.



\section{Model grid and comparison strategy}
\label{sec:models}

\subsection{Methodology: model--data comparison}
\label{subsec:method}

Our aim is not to derive a formal best fit for each star, but to identify evolutionary solutions that remain physically plausible within a small set of related single-star models.
For each object, we first search the considered tracks for epochs whose HR diagram positions are closest to the observed $(L,T_\ast)$.
We then retain only those candidates whose evolutionary phase is consistent with the observed subtype and surface-composition behaviour, and finally evaluate them using the wind-scaling diagnostics defined in Sect.~\ref{subsec:vesc}.

In this comparison, luminosity is given higher weight than temperature when identifying HR diagram candidates.
This is because the atmosphere-based quantity $T_\ast$ is defined at the inner boundary of the expanding atmosphere and does not map uniquely onto the temperature variable used in stellar-evolution calculations; in addition, the inferred WN temperature scale can depend on the adopted hydrogen content of the atmosphere model \citep{Hamann2006,Crowther2007,Sander2012,Hamann2019}.
We therefore treat the temperature agreement as a secondary constraint and discuss this systematic issue separately in Sect.~\ref{sec:discussion}.

When more than one candidate remains viable, we retain two representative solutions (Model~A and Model~B) to bracket the main residual degeneracy in evolutionary timing or track family.
These labels are assigned separately for each star and do not denote one fixed pair of model families throughout the paper.
If only one candidate remains, we retain a single model point; if none satisfies the combined constraints, we report that outcome explicitly.


\subsection{Adopted evolutionary framework}
\label{subsec:evo_framework}

We compare the observed stars with a small set of closely related MESA (version r12115, developed by \citet{Paxton2011,Paxton2013,Paxton2015,Paxton2018,Paxton2019}) evolutionary sequences based on the \citet{Li2023} framework at approximately solar metallicity, which adopts a turbulence-motivated $k$--$\omega$ treatment of convection and convective-boundary mixing.
The initial composition is $Z=0.02$, $X=0.70$, and $Y=0.28$. The main low-mass grid emphasized in this work covers initial masses from $16$ to $40\,M_\odot$. Unless otherwise stated, the rotating models start on the zero-age main sequence (ZAMS) with $v_{\rm ini}/v_{\rm crit}=0.4$, and all models are computed up to the end of core carbon burning.
This framework is used as the common backbone of the analysis, while different track families probe the sensitivity to the adopted wind treatment and, where relevant, to rotation.

We adopt the MESA ``Dutch'' wind scheme with an overall scaling factor $\eta_{\rm Dutch}=1.0$, except in models where alternative enhanced-wind prescriptions are explicitly explored. The Schwarzschild criterion is replaced by the Ledoux criterion for determining convective boundaries. Convective core overshooting is treated self-consistently using the $k-\omega$ model. We also account for overshooting in convective shells and convective envelopes. For simplicity, overshooting in these regions is treated using the exponentially decaying diffusive overshooting prescription implemented in MESA, which is based on the standard mixing-length theory (MLT). The criteria used to distinguish the different WR subtypes are based on the general definitions adopted by \citet{Li2023}.

\subsection{Mass-loss prescription}
\label{subsec:mloss}

The compared track families share the same general MESA backbone, but differ in the adopted wind prescription.
The model labels used below indicate the mass-loss prescriptions adopted in the red-supergiant (RSG) and WR phases: the first part of the label refers to the RSG wind, while the second part refers to the WR wind.
Hereafter, S99 denotes the RSG prescription of \citet{1999A&A...342..131S}, Y23 denotes the empirical RSG prescription of \citet{Yang2023}, S20 denotes the WR prescription of \citet{Sander2020}, and HS19 denotes the empirical WR-wind calibration based on the Galactic WN and WC atmosphere analyses of \citet{Hamann2019} and \citet{Sander2019}.
For example, S99+S20 denotes the S99 prescription during the RSG phase and the S20 prescription during the WR phase, whereas Y23+S20 uses the Y23 RSG prescription with the same WR-phase prescription.

For the RSG phase, we consider two prescriptions.
The S99 prescription follows \citet{1999A&A...342..131S},
\begin{equation}
\log \dot{M} = 2.1\,\log(L/L_\odot)-14.5 ,
\end{equation}
where $\dot{M}$ is in $M_\odot\,{\rm yr^{-1}}$.
The Y23 prescription follows the empirical RSG mass-loss relation of \citet{Yang2023},
\begin{equation}
\log \dot{M} =
0.45\,x^3 - 5.26\,x^2 + 20.93\,x - 34.56 ,
\end{equation}
where $x=\log(L/L_\odot)$.

For the WR phase, the reference tracks adopt the S20 prescription of \citet{Sander2020}, which expresses the WR mass-loss rate as a function of the electron-scattering Eddington factor and metallicity:
\begin{equation}
\log \dot{M}
=
a\,\log[-\log(1-\Gamma_{\rm e})]
-\log(2)\left(\frac{\Gamma_{\rm e,b}}{\Gamma_{\rm e}}\right)^c
+\log \dot{M}_{\rm off}.
\end{equation}
Here, $a=2.932$, $c=-0.44\,\log(Z/Z_\odot)+9.15$, and $\log \dot{M}_{\rm off}=0.23\,\log(Z/Z_\odot)-2.61$. $Z$ is the metallicity, $Z_\odot=0.014$ is the solar metallicity adopted in \citet{Sander2020}.  The parameter $\Gamma_{\rm e}$ depends primarily on the stellar luminosity-to-mass ratio ($L/M$), and is defined as
\begin{equation}
	\Gamma_{\rm e} = \frac{\sigma_{\rm e}}{4cm_{\rm H}G} q_{\rm ion} \frac{L}{M} =10^{-4.51} q_{\rm ion} \frac{L/L_\odot}{M/M_\odot},
\end{equation}
$q_{\rm ion} \approx 0.5$ for WR stars, $\Gamma_{\rm e,b}=-0.324\,\log(Z/Z_\odot)+0.244$,
For selected WNC and WC comparisons, we also test the HS19 empirical WR-wind calibration based on the Galactic WR atmosphere analyses of \citet{Hamann2019} and \citet{Sander2019}.

When the HS19 empirical calibration is used, the adopted WR mass-loss rate is switched according to the instantaneous surface composition of the model.
For H-rich WN stars (WNL), we use
\begin{equation}
\log \dot{M} = 0.55\,\bigl[\log(L/L_\odot)-5.22\bigr]-5.01 .
\end{equation}
For H-free WN stars (WNE) and WNC stars, we use
\begin{equation}
\log \dot{M} = 0.89\,\bigl[\log(L/L_\odot)-4.92\bigr]-5.12 .
\end{equation}
For WC/WO stars, we use
\begin{equation}
\log \dot{M} = -8.68 + 0.71\,\log(L/L_\odot) - 0.74\,\log Y ,
\label{eq:HS19_WC}
\end{equation}
where $Y$ is the surface He mass fraction.
The transitional WNC phase is assigned according to the instantaneous surface composition and is followed on the WN-like or WC-like side of the transition as appropriate.

Throughout the paper, we distinguish between the observational spectral subtype assigned to each target and the evolutionary phase assigned to a model timestep.
The former is taken from the adopted observational catalogue, whereas the latter follows the surface-composition criteria used in Paper~II.
This distinction is important for objects such as WR\,120, which has a late WN spectral subtype but is classified as WN7/WNE-w in the adopted sample because of its H-poor atmosphere.


\subsection{Model quantities used in the comparison}
\label{subsec:model_quantities}

From each evolutionary track, we extract at each timestep the stellar luminosity, temperature, current mass, radius, and surface abundances. Throughout this paper, $X_i$ denotes the surface mass fraction of element $i$, and $Y_{\rm c}$ denotes the central He mass fraction. Values of $X_i$ and $Y_{\rm c}$ listed in the tables are given in percent, whereas values quoted in the text are given as fractions unless explicitly followed by a percent sign. For WNC stars, C/N always refers to the mass ratio $X_{\rm C}/X_{\rm N}$; its logarithmic form is $\log(X_{\rm C}/X_{\rm N})$.
These quantities are used to identify HR diagram candidates, to test subtype/composition consistency, and to compute the wind-scaling diagnostics for the selected model points.
This procedure ensures that the wind diagnostics are evaluated using self-consistent $(M,R)$ values from the evolutionary models.

A global comparison between the observational sample and representative evolutionary tracks is presented at the beginning of Sect.~\ref{sec:results}.


\section{Results}
\label{sec:results}

\subsection{Global HR diagram comparison}
\label{subsec:global_hrd}

\begin{table*}[t]
\centering
\caption{Parameters of the low-luminosity Galactic WR stars.}
\label{tab:Table_LowLWR}
\small
\setlength{\tabcolsep}{5pt}
\begin{tabular}{lllcccccccccccc}
\hline \hline
WR & Subtype & Spectral & Binary & $\log L$ & $T_\ast$ & $R$ & $M$ & $\log\dot{M}$ 
   & $X_{\rm H}$ & $X_{\rm He}$ & $\log R_{\rm t}$ & $v_\infty$ & $\eta_{\rm wind}$ & Ref. \\
   &  & subtype & status & $(L_\odot)$ & (kK) & ($R_\odot$) & ($M_\odot$) & ($M_\odot\,\mathrm{yr}^{-1}$) & \multicolumn{2}{c}{(mass frac. \%)} & ($R_\odot$) & (km\,s$^{-1}$) &  &  \\
   (1) & (2) & (3) & (4) & (5) & (6) & (7) & (8) & (9) & (10) & (11) & (12) & (13) & (14) & (15) \\
\midrule
\multicolumn{15}{c}{$5.2 < \log L/ L_\odot \leq 5.4$} \\
123    & WNL$^{(a)}$ & WN8  & Yes& 5.28 & 44.7 & 6.97 & 12.0 & -4.60 & 0    &   & 0.7 & 970  & 6.7 & 1, 3 \\
82     & WNL & WN7(h)  & No & 5.26 & 56.2 & 4.24 & 11.0 & -4.80 & 20 &   & 0.7 & 1100 & 4.9 & 1 \\
55     & WNL$^{(a)}$ & WN7  & No & 5.40 & 56.2 & 5.23 & 14.0 & -4.70 & 0    &   & 0.8 & 1200 & 4.7 & 1 \\
84     & WNL$^{(a)}$ & WN7  & No & 5.36 & 50.1 & 6.30 & 13.0 & -4.80 & 0    &   & 0.9 & 1100 & 3.6 & 1 \\
85     & WNL & WN6h-w  & Yes& 5.38 & 50.1 & 6.46 & 13.0 & -5.00 & 40 &   & 1.1 & 1400 & 3.1 & 1, 3 \\
49     & WNE-w$^{(b)}$ & WN5h-w  & Yes& 5.40 & 56.2 & 5.20 & 14.0 & -5.00 & 25 &   & 1.0 & 1450 & 2.8 & 1 \\
36     & WNE-s & WN5-s & No & 5.30 & 89.1 & 1.79 & 12.0 & -4.30 & 0    &   & 0.2 & 1900 & 23.6& 1, 3 \\
128    & WNE-w$^{(b)}$ & WN4h-w  & Yes& 5.22 & 70.8 & 2.69 & 11.0 & -5.40 & 16 &   & 1.1 & 2050 & 2.6 & 1, 3 \\
129    & WNE-w & WN4-w & No & 5.40 & 63.1 & 4.17 & 14.0 & -5.00 & 0    &   & 0.9 & 1320 & 2.4 & 1 \\
7      & WNE-s & WN4-s & No & 5.36 & 112.2& 1.26 & 13.0 & -4.80 & 0    &   & 0.3 & 1600 & 5.9 & 1 \\
81     & WCL & WC9     & No & 5.26 & 45.0 & 7.08 & 11.1 & -4.62 & 0 & 55 & 0.8 & 1600 & 10.4& 2 \\
69     & WCL & WC9d    & Yes& 5.33 & 40.0 & 9.77 & 12.1 & -4.87 & 0 & 55 & 1.0 & 1089 & 3.3 & 2, 3 \\
80     & WCL & WC9d    & No & 5.24 & 45.0 & 6.89 & 10.8 & -4.79 & 0 & 55 & 0.9 & 1600 & 7.5 & 2 \\
95     & WCL & WC9d    & No & 5.23 & 45.0 & 6.86 & 10.7 & -4.71 & 0 & 55 & 0.9 & 1900 & 10.8& 2 \\
106    & WCL & WC9d    & No & 5.23 & 45.0 & 6.81 & 10.6 & -4.80 & 0 & 55 & 0.8 & 1100 & 5.1 & 2 \\
117    & WCL & WC9d    & No & 5.36 & 56.0 & 5.12 & 12.5 & -4.44 & 0 & 55 & 0.6 & 2000 & 15.5& 2 \\
135    & WCL & WC8     & No & 5.40 & 63.0 & 4.24 & 13.6 & -4.73 & 0 & 75 & 0.6 & 1343 & 4.9 & 2 \\
56     & WCL & WC7     & No & 5.33 & 71.0 & 3.09 & 12.1 & -4.76 & 0 & 55 & 0.6 & 2009 & 7.9 & 2 \\
64     & WCL & WC7     & No & 5.27 & 71.0 & 2.87 & 11.2 & -4.88 & 0 & 55 & 0.6 & 1700 & 5.9 & 2 \\
27     & WCE & WC6     & Yes& 5.28 & 79.0 & 2.35 & 11.3 & -4.78 & 0 & 55 & 0.5 & 2100 & 9.0 & 2, 3 \\
132    & WCE & WC6     & Yes& 5.35 & 71.0 & 3.15 & 12.4 & -4.67 & 0 & 55 & 0.6 & 2400 & 11.3& 2, 3 \\
111    & WCE & WC5     & No & 5.39 & 89.0 & 2.10 & 13.0 & -4.64 & 0 & 55 & 0.4 & 2398 & 11.0& 2 \\
114    & WCE & WC5     & Yes& 5.39 & 79.0 & 2.68 & 13.1 & -4.51 & 0 & 55 & 0.5 & 3200 & 19.7& 2, 3 \\
38     & WCE & WC4     & No & 5.21 & 126.0& 0.85 & 10.4 & -4.66 & 0 & 55 & 0.1 & 3200 & 21.6& 2 \\
\hline
\multicolumn{15}{c}{$\log L/ L_\odot \leq 5.2$} \\
120    & WNL$^{(a)}$ & WN7   & No & 4.92 & 50.1 & 3.78 & 7.0  & -4.90 & 0    &   & 0.8 & 1225 & 8.9 & 1 \\
67     & WNE-w & WN6-w & No & 5.11 & 56.2 & 3.73 & 9.0  & -4.80 & 0    &   & 0.8 & 1500 & 8.7 & 1 \\
71     & WNE-w & WN6-w & Yes& 5.06 & 56.2 & 3.56 & 9.0  & -5.10 & -    &   & 0.9 & 1200 & 3.7 & 1, 3 \\
61     & WNE-w & WN5-w & No & 5.03 & 63.1 & 2.75 & 9.0  & -5.00 & 0    &   & 0.7 & 1400 & 6.8 & 1 \\
8      & WNC & WN6/WC4 & No & 5.10 & 48.0 & 5.00 &      & -4.20 & 0 &      &     & 1590 &  39.4& 4 \\
58     & WNC & WN4/WCE & No & 4.95 & 79.0 & 1.61 & 8.4  & -4.95 & 0 & 98 & 0.5 & 1600 &  9.9 & 2 \\
121--16& WNC & WN7o/WC & No & 4.88 & 47.0 & 4.14 & 7.1  & -4.97 & 0 & 98 & 0.8 & 1000 &  7.0 & 5 \\
92     & WCL & WC9     & No & 4.95 & 45.0 & 4.97 & 7.7  & -5.00 & 0 & 55 & 0.8 & 1121 & 6.1 & 2 \\
119    & WCL & WC9d    & No & 4.70 & 45.0 & 3.70 & 5.8  & -5.13 & 0 & 55 & 0.8 & 1300 & 9.6 & 2 \\
121    & WCL & WC9d    & No & 5.16 & 45.0 & 6.35 & 9.9  & -4.85 & 0 & 55 & 0.8 & 1100 & 5.3 & 2 \\
52     & WCE & WC4     & No & 5.07 & 112.0& 0.92 & 8.5  & -4.75 & 0 & 35 & 0.2 & 3225 & 24.1& 2 \\
144    & WCE & WC4     & No & 5.20 & 112.0& 1.06 & 9.9  & -4.62 & 0 & 35 & 0.2 & 3500 & 26.4& 2 \\
\hline
\end{tabular}
\tablefoot{
\tablefoottext{a}{These stars were classified as WNE-w stars by \citet{Hamann2019}. We treat them here as WNL stars because their spectral subtypes are late types (WN7-WN9), despite the fact that their Potsdam Wolf-Rayet (PoWR) atmosphere models indicate no detectable hydrogen in these stars.}
\tablefoottext{b}{These stars were classified as WNL stars by \citet{Hamann2019}. We consider them here as WNE-w stars, despite their relatively high surface H mass fractions ($X_{\rm H}>0.1$).} 
\tablebib{1~\citet{Hamann2019}; 2~\citet{Sander2019}; 3~\citet{2001NewAR..45..135V}; 4~\citet{1995A&A...304..269C}; 5~\citet{zhang2020}.}
}
\end{table*}

To compare with the evolutionary tracks for the models with enhanced mass-loss rate during the RSG phase, we selected low-luminosity ($\log L/L_\odot \leq 5.4$) single WR stars from \citet{Hamann2019} and \citet{Sander2019}, supplemented by WR~8 and WR~121--16 from \citet{1995A&A...304..269C} and \citet{zhang2020}. 
Some of their physical parameters are listed in Table~\ref{tab:Table_LowLWR}. 

Table~\ref{tab:Table_LowLWR}, Col. (1) lists the IDs of low-luminosity WR star samples.  
Col. (2) gives the WR subtypes classified according to their spectral types. In general, WN stars with subtypes ranging from WN2 to WN5 are classified as early-type WN stars (WNE), while those from WN7 to WN11 are classified as late-type WN stars (WNL); WN6 stars can be either early or late type \citep{Crowther2007,2024arXiv241004436S}; 
The spectral types listed in Col. (3) are taken basically from the WR catalog. For WNE stars, we append “-w” or “-s” to the classifications to indicate weak or strong emission lines, respectively.
The binary status for the WR stars, presented in Col. (4), is taken from the WR catalog \citep{2001NewAR..45..135V};
The physical parameters which can be compared with evolutionary models given by references are listed in Table~\ref{tab:Table_LowLWR} Col. (5) to (14);
The references for the selected WR stars are listed in Col. (15). The WN and WC samples are adopted from \citet{Hamann2019} and \citet{Sander2019}, respectively. Two additional WNC stars with luminosities lower than $\log L/L_\odot = 5.2$ are taken from \citet{1995A&A...304..269C} and \citet{zhang2020}.

These samples are ordered according to their spectral subtypes, ranging from WN8 to WC4. The lowest luminosity WR stars with $\log L/L_\odot \leq 5.2$ are listed separately on the bottom of the Table. 
In this table, we also list the WR star samples that are members of binary systems. Binary evolution is required to explain the lower luminosity WC stars \citep{Sander2019}. However, only a small fraction of the samples in the table are members of binary systems. In particular, among the extremely low-luminosity WR stars ($\log L/L_\odot \leq 5.2$), only one sample is identified as a binary. 
While binary identification completeness cannot be fully assessed for such a small and heterogeneous sample, the predominance of apparently single objects among the faintest WR stars suggests that single-star mass loss, particularly enhanced winds during the RSG phase, warrants serious consideration as a contributing channel.

\begin{figure*}[ht!]
\centering
\includegraphics[width=\textwidth]{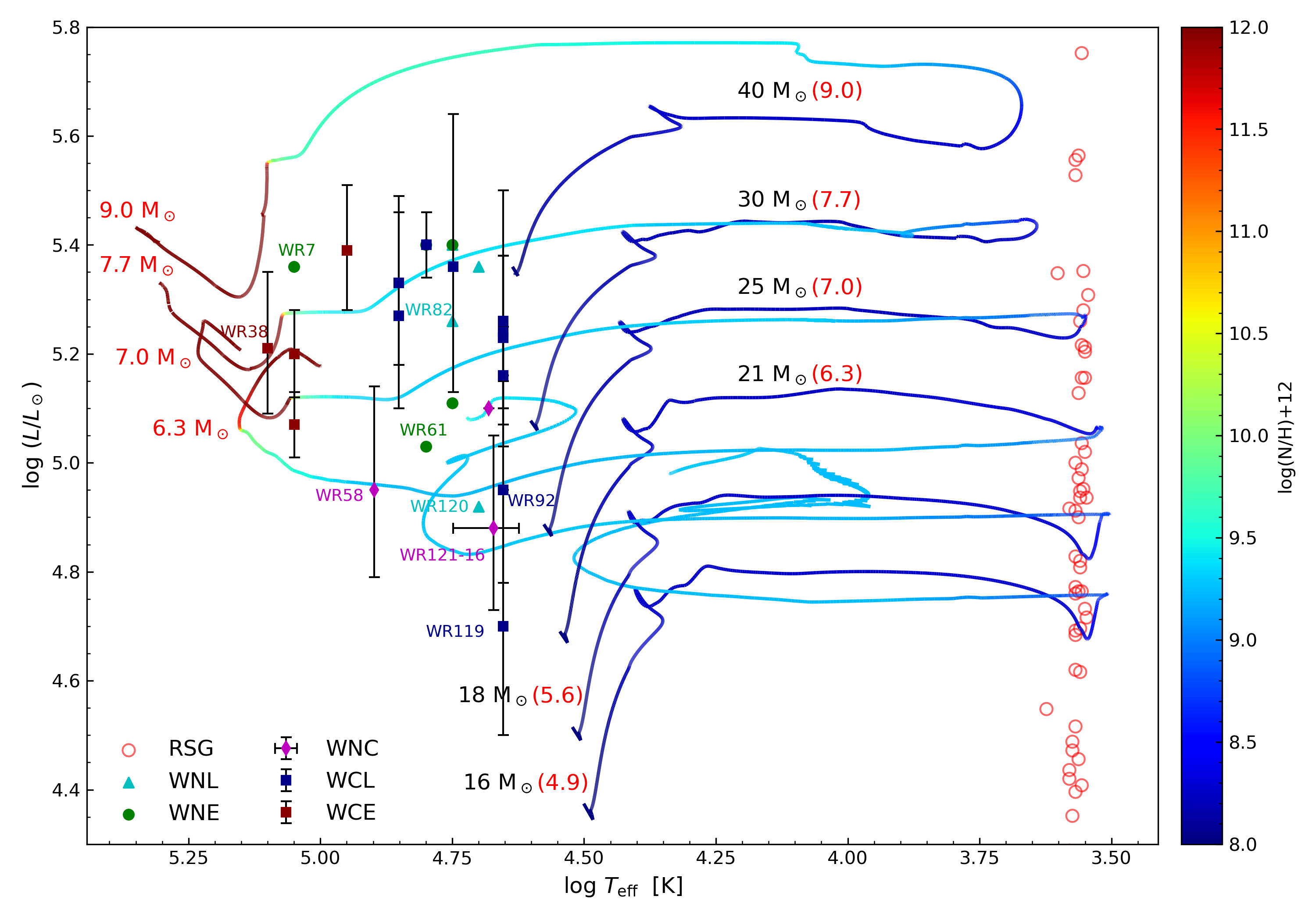}
\caption{
HR diagram comparison for the low-luminosity WR sample. Observed single stars are shown by subtype, with error bars in $\log L$ and $\log T_{\rm eff}$. 
Representative rotating single-star evolutionary tracks with initial masses from 16 to 40 $M_\odot$ are overplotted. The values shown in red indicate the final masses of the corresponding models. 
Mass loss during the RSG and WR phases is treated using the S99 and S20 prescriptions, respectively.
Red open circles denote RSGs in the Milky Way \citep{2005ApJ...628..973L}. The color scale indicates the surface number abundance, $\rm \log(N/H) + 12$, along the tracks.
}
\label{fig:HRD_NH}
\end{figure*}

Figure~\ref{fig:HRD_NH} summarizes the updated HR diagram comparison between the observed single WR sample and the evolutionary tracks calculated by MESA with the $k-\omega$ model. 
The minimum initial mass required for a star to evolve into a WR
star decreases to about $18\,M_\odot$ when the S99 mass-loss
prescription is adopted during the RSG phase.
As with the results of \citet{1999A&A...342..131S}, the enhanced mass-loss rates efficiently strip the stellar envelopes. This provides a simple explanation of the blue extension of the loop, and the models may even evolve into WR stars once the surface H mass fraction decreases to 0.4 at the stage of central He-exhaustion. The reduction of the surface H mass fraction makes the models hotter because of the higher mean molecular weight. The luminosities decrease slightly as the models evolve into the WR stage. As the mass loss prescription changes from \citet{Vink2001} to S20, the more efficient stellar wind rapidly reduces the stellar radii.

The grid of evolutionary tracks from 16 to $40\,M_\odot$ can reproduce the positions of most Galactic low-luminosity WR stars. 
In the upper luminosity range ($5.2 < \log L/ L_\odot \leq 5.4$), models with initial mass between 25 and $40\,M_\odot$ match the observational samples well. 
The color scale changes from cyan to green, roughly corresponding to a decrease in $X_{\rm H}$ from 0.5 to 0. This broad range in the HR diagram traces the WN phase evolving from WNL to WNE. The WNE phase occupies only a narrow region near the turning point, where the color scale changes to red as hydrogen becomes depleted. 
The red colour scale marks models evolving from the WNC to the WC stage. The luminosity drops substantially at this point and the tracks turn to hotter, more luminous configurations after the end of central He-burning. 
Like most evolutionary grids of massive stars, the WNE and WC stars predicted by the evolutionary models can only reproduce the hotter WR samples. The “temperature problem” remains unsolved even after adopting the new mass-loss prescription. A possible key to resolving this problem is that the stars may undergo significant expansion during the transition from the WNE stage to the WCL stage.

In the lower luminosity range ($\log L/ L_\odot \leq 5.2$), the situation becomes even less favourable once the subtype constraints are taken into account. Models with lower initial masses reach the main sequence band once the H mass fraction in the envelope decreases below about 0.45. 
The $16\,M_\odot$ model reaches this region only near the end of central He burning, then moves back to the red side of the HR diagram and eventually ends its evolution as a blue supergiant (BSG) or luminous blue variable (LBV). 
For the $18\,M_\odot$ model, the luminosity increases noticeably in the WN region after central helium exhaustion, and the model ultimately ends as a WNL star with a relatively thin hydrogen envelope.
With increasing initial mass, the stars are more likely to evolve into the WNC phase after most of the H-rich envelope has been stripped away, and eventually end their evolution as WNC stars. Even considering the large luminosity uncertainties of the observed WNC sample, the luminosities predicted by the evolutionary models remain slightly higher than the observed values.
For WC stars, the situation is similar to that found in the upper luminosity range, where the “temperature problem” still cannot be resolved. Since low-mass WR stars enter the WR phase only during the late stage of core He burning, they do not have sufficient time to evolve into the WC phase, and therefore cannot explain the WC stars with luminosities lower than $\log L/L_\odot = 5.0$.

The H-rich WN object WR\,82 is close to the selected H-rich WN phases, whereas the HR diagram position of WR\,120 can be approached even though its adopted WNE-w classification makes the surface-composition constraint more restrictive. 
The WNE stars can be reached after substantial envelope removal. 
The WNC stars remain more restrictive because they require a narrow transitional surface composition. 
For the WC stars, the WCL objects WR\,92 and WR\,119 are approached in luminosity by the HS19-based tracks but remain cooler than the selected model points, whereas the WCE object WR\,38 is much better matched in the HR diagram. 
Thus, the updated comparison does not support a single uniform conclusion for all low-luminosity WR stars.

We therefore present the subtype-by-subtype comparison below, beginning with the WN reference objects and then turning to the WNC and WC stars, where the strongest constraints arise from the simultaneous requirements of HR diagram location, surface composition, and WR-like wind properties.


\subsection{WN stars}
\label{subsec:results_wn}

\subsubsection{WNL stars}
\begin{figure*}[ht!]
\centering
	\includegraphics[width=0.495\textwidth]{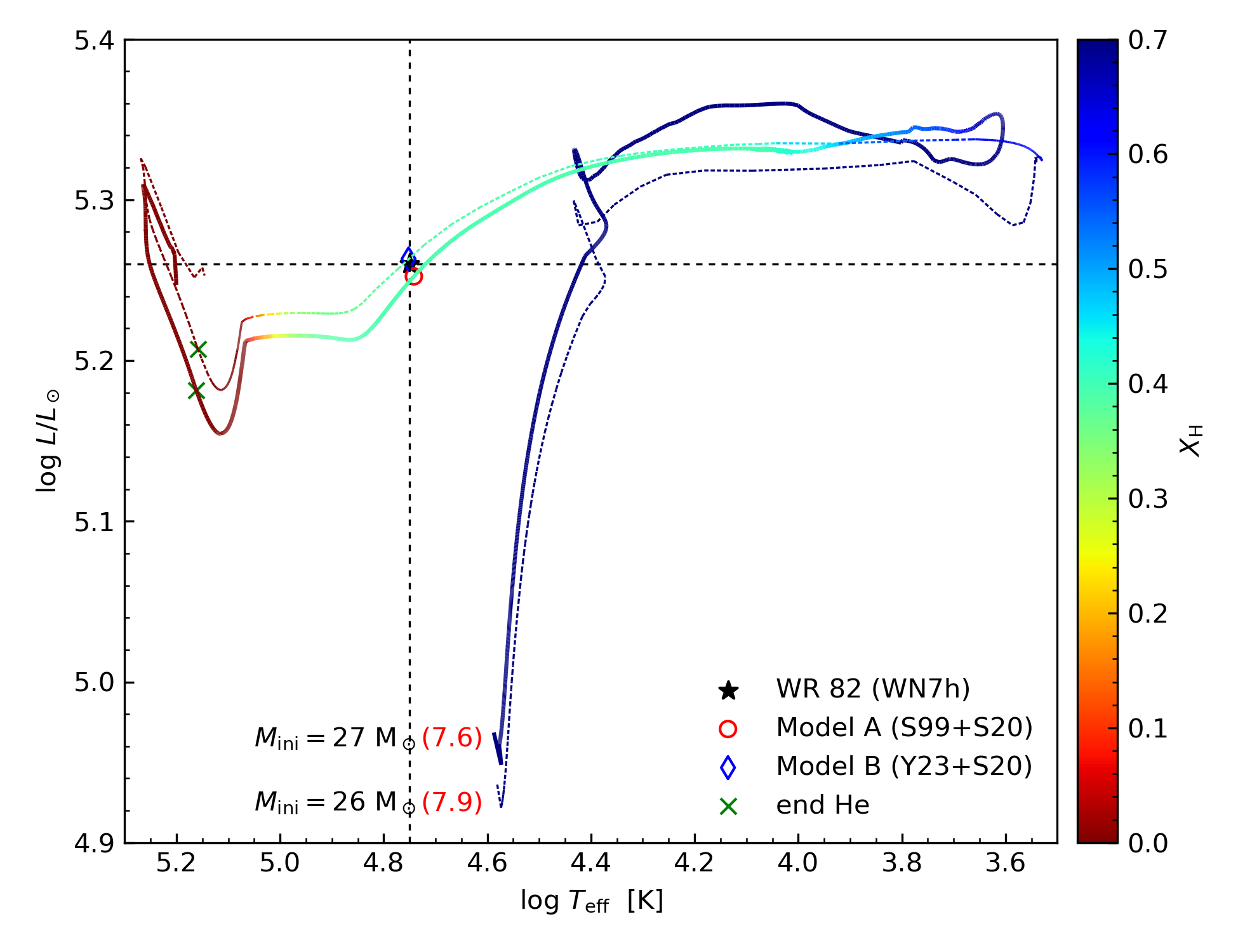}
	\hfill
	\includegraphics[width=0.495\textwidth]{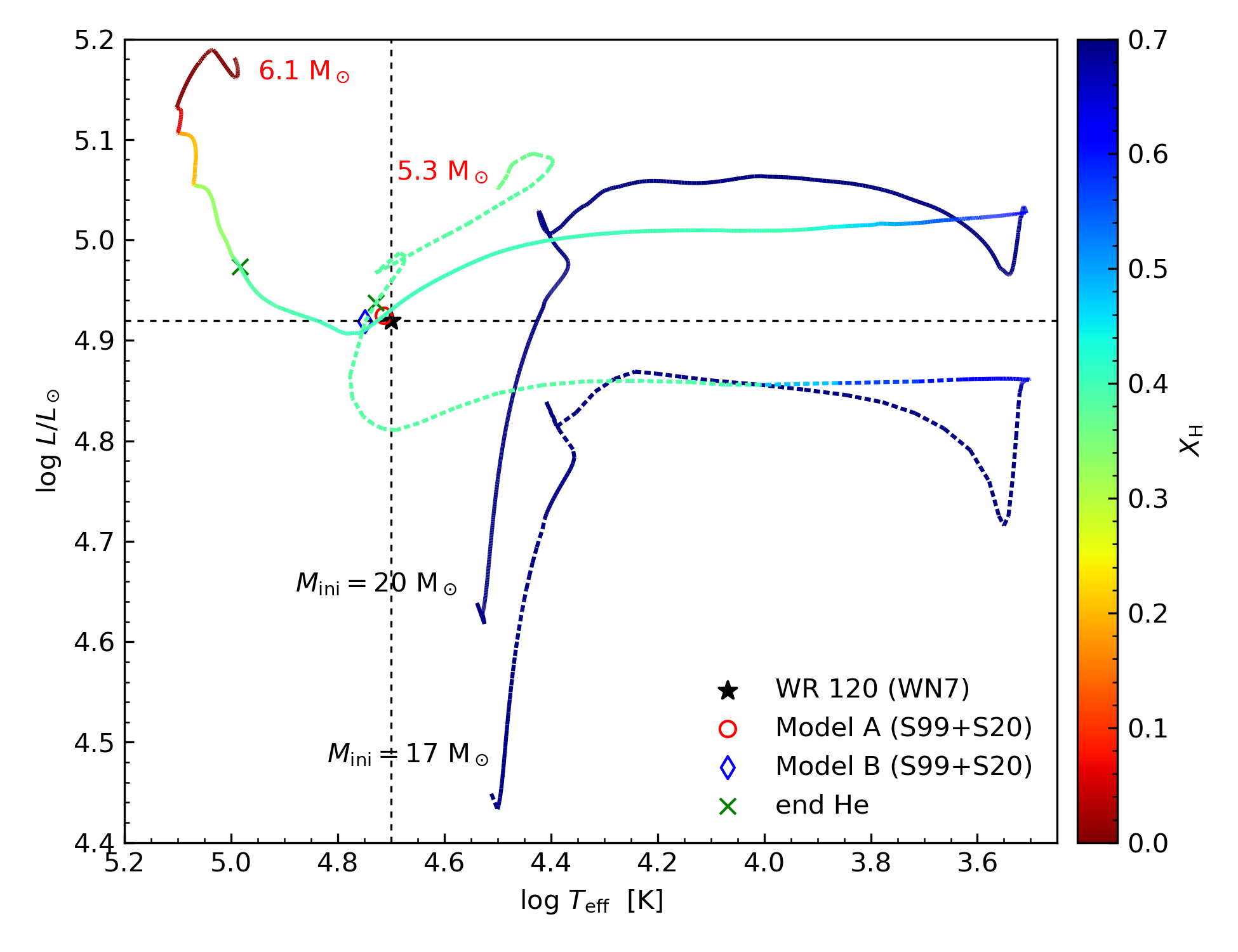}
\caption{
HR diagram comparisons between evolutionary models and two representative WNL stars, WR\,82 and WR\,120. WR\,82 is classified as WN7(h), while WR\,120 is classified as WN7 (WNE-w) in the adopted observational sample.
The evolutionary tracks are labelled by their initial mass, and the values shown in red indicate the final masses.
The solid and dashed lines represent models computed with the S99+S20 and Y23+S20 mass-loss prescriptions, respectively.
The black star indicate the adopted observed positions. Highlighted symbols mark the representative best-fitting Model~A and Model~B points, while the green crosses indicate the end of core He-burning.
The colour scale indicates the surface H mass fraction, \(X_{\rm H}\).
}
\label{WNL}
\end{figure*}

Comparisons for two selected WNL stars are shown in Figure~\ref{WNL}.
The two targets are interpreted according to the adopted observational classifications: WR\,82 is a H-rich WN7(h) star, whereas WR\,120 has a late WN spectral subtype but is classified as WN7 (WNE-w) by the PoWR atmosphere analysis of \citet{Hamann2019}, which indicates no detectable surface H.

The parameters of the low-luminosity WN stars are listed in Table~\ref{tab:Table_WN}. Some of these physical parameters are also included in Table~\ref{tab:Table_LowLWR}, allowing a direct comparison with the evolutionary models. 
Col. (3) lists the mass-loss prescriptions adopted during the RSG and WR phases. Columns (8) and (9) give the initial mass and the mass of the H-rich envelope, respectively. The central He mass fraction, $Y_{\rm c}$, is listed in Col. (12). Col. (13), $v_{\rm eq}$, gives the equatorial surface rotational velocity. The last column gives the ages of the best-fitting models, which can be used to estimate the ages of the corresponding observed stars. 

\begin{table*}[t]
\centering
\caption{
Observed parameters and representative Model~A and Model~B solutions for the selected low-luminosity WN stars.
}
\label{tab:Table_WN}
\small
\setlength{\tabcolsep}{5pt}
\begin{tabular}{lllccccccccccccr}
\hline \hline
Model & Subtype & MLR & $v/v_{\rm crit}$ & $\log L$ & $T$ & $R$ & $M$ & $M_{\rm ini}$ & $M_{\rm env}$
& $X_{\rm H}$ & $X_{\rm He}$ & $Y_{\rm c}$ & $v_{\rm eq}$ & $\log\dot{M}$ & Age \\
 & & & & $(L_\odot)$ & (kK) & ($R_\odot$) & ($M_\odot$) & ($M_\odot$) & ($M_\odot$)
& \multicolumn{3}{c}{(mass frac. \%)} & (km\,s$^{-1}$) & ($M_\odot\,{\rm yr}^{-1}$) & (Myr) \\
   (1) & (2) & (3) & (4) & (5) & (6) & (7) & (8) & (9) & (10) & (11) & (12) & (13) & (14) & (15) & (16) \\
\hline
WR~82  & WN7(h)  &  &  & 5.26 & 56.2 & 4.24 & 11.0 &  &  & 20 &  &  &  & -4.80 &  \\
A   & WNL & S99+S20 & 0.4 & 5.25 & 55.2 & 4.63 & 10.30 & 27 & 0.95 & 39 & 59 & 43 & 3.36 & -5.13 & 6.74 \\
B   & WNL & Y23+S20 & 0.4 & 5.26 & 56.5 & 4.47 & 10.14 & 26 & 0.86 & 38 & 60 & 27 & 3.41 & -5.07 & 7.04 \\
WR~120 & WN7     &  &  & 4.92 & 50.1 & 3.78 & 7.0  &  &  & 0    &   &  &  & -4.90 &  \\
A   & WNL & S99+S20 & 0.4 & 4.92 & 51.7 & 3.62 & 6.82 & 20 & 0.69 & 40 & 58 & 37 & 4.44 & -5.97 & 9.08 \\
B   & WNL & S99+S20 & 0.4 & 4.92 & 56.1 & 3.06 & 5.54 & 17 & 0.30 & 38 & 60 & 0 & 2.99 & -5.40 & 11.47 \\
\hline
WR~7   & WN4-s &    &  & 5.36 & 112.2 & 1.26 & 13.0  &  &  & 0 &  &  &  & -4.80 &  \\
A & WNE-s & S99+S20 & 0.4 & 5.36 & 121.0 & 1.09 & 11.93 & 33 & 0 & 0 & 98 & 46 & 230 & -4.90 & 5.72 \\
B & WNE-s & S99+S20 & 0.6 & 5.36 & 121.3 & 1.08 & 12.17 & 33 & 0 & 0 & 98 & 56 & 297 & -4.90 & 5.86 \\
WR~61  & WN5-w &    &  & 5.03 & 63.1  & 2.75 & 9.0  &  &  & 0    &   &  &  & -5.00 &  \\
A-1 & WNL & S99+S20 & 0.4 & 5.03 & 63.9  & 2.67 & 7.98 & 23 & 0.55 & 40 & 58 & 36 & 4.11 & -5.66 & 7.86 \\
A-2 & WNE-s & S99+S20 & 0.4 & 5.04 & 112.8 & 0.86 & 7.43 & 23 & 0.001 & 1 & 97 & 12 & 243 & -5.37 & 8.05 \\
B & WNE-s & S99+S20 & 0.2 & 5.06 & 115.3 & 0.85 & 7.45 & 23 & 0 & 0 & 98 & 8 & 112 & -5.33 & 7.98 \\
\hline
\end{tabular}
\tablefoot{
Model~A and Model~B denote representative evolutionary points selected from the corresponding track families. 
MLR denotes the adopted mass-loss-rate prescription.
The observed reference parameters and their literature sources are given in Table~\ref{tab:Table_LowLWR}. 
}
\end{table*}

\vspace{0.2cm}
\noindent \textbf{WR\,82}: both selected tracks pass very close to the observed position in the HR diagram.
The representative Model~A and Model~B points with different mass-loss prescriptions in RSG phase, are nearly coincident with the observational cross.
The colour scale also shows that the selected evolutionary phases retain a substantial surface H mass fraction, as expected for a H-rich WN7(h) interpretation.
Both Model~A and Model~B correspond to the very beginning of the WNL phase and with relatively high surface H mass fractions of $X_{\rm H}\simeq 0.4$ (see Table~\ref{tab:Table_WN}), which is approximately twice the value inferred from the observations ($X_{\rm H}=0.2$). 

Although the envelope still retains a relatively high hydrogen abundance, the mass of the H-rich envelope (defined as the total stellar mass minus the He-core mass) is already below $1\,M_\odot$. The total masses predicted by the best-fitting models are somewhat lower than the current stellar masses derived from the mass–luminosity relation for homogeneous helium stars adopted by \citet{Hamann2019}. Both models predict slightly larger stellar radii, and very slow stellar rotation, as the surface layers are far away from the rapidly rotating helium core. At the model point corresponding to the observed luminosity, the mass-loss rate predicted by the S20 prescription is slightly lower than the observed value. The best-fitting evolutionary models predict that WR\,82 has an age of approximately 7 Myr.

The $26\,M_\odot$ model with the Y23 mass-loss prescription during the RSG phase produces a best-fitting model very close to Model~A. Owing to its lower mass-loss rate during the RSG phase, Model~B remains in this stage for a longer period before entering the WR phase. Consequently, it reaches the observed position with a lower central He mass fraction and a greater age than Model~A.

\vspace{0.2cm}
\noindent \textbf{WR~120} is a WN7 star in the Galactic WR catalogue. \citet{2023A&A...674A..88D} measured a peak-to-peak radial-velocity amplitude of $\Delta\,{\rm RV}=40.5\,{\rm km\,s^{-1}}$ from seven epochs spanning 385 days, and classified the object as a single star. The luminosity of WR\,120, $\log(L/L_\odot)=4.92$, is much lower than that of typical WNL stars and is therefore not expected to be reproduced by standard single-star evolutionary tracks without enhanced mass loss. To approach the observed luminosity and temperature, we selected two low-initial-mass models computed with the S99+S20 prescription.

Although the two best-fitting models correspond to different evolutionary channels and timings, both provide a good match to the observed HR diagram position. Model~A reaches the observed position after crossing the main-sequence band and continues to evolve bluewards as stellar winds peel off the H-rich envelope. It ends as a H-free WNE star after central carbon exhaustion. Model~B, with an even lower initial mass, provides another possible channel to reach the same region. Similar to the $18\,M_\odot$ model in Fig.~\ref{fig:HRD_NH}, Model~B moves rapidly toward higher luminosities in the HR diagram as central helium exhaustion is approached.

The best-fitting Model~B is located almost exactly at the position marked by the green cross, which denotes the end of core He-burning. Moreover, this model remains in the WNL phase until the end of core C-burning. Although Model~B has a surface H mass fraction similar to that of Model~A, $X_{\rm H}\simeq 0.4$, and both correspond to the early WNL phase, Model~B possesses a significantly less massive H-rich envelope. This indicates that Model~B has undergone more extensive envelope stripping and is therefore at a more advanced evolutionary stage. Consequently, its age is also substantially greater than that of Model~A.

The result suggests that massive stars with initial masses below $19\,M_\odot$ may evolve into WR stars if they experience sufficiently enhanced mass loss during the RSG phase, and may eventually end their lives as WNL stars. As progenitors of core-collapse supernovae, these stars retain only very thin hydrogen envelopes, with $M_{\rm env}<0.5\,M_\odot$. Consequently, they are expected to explode as Type IIb or possibly Type Ibn supernovae, depending on the properties of the remaining circumstellar material (CSM).

Compared with stripped-envelope Type IIb supernova progenitors from binary models \citep{2022ApJS..262...26L}, our models have higher effective temperatures and larger helium-core masses. Some models end in the region between WNL stars and RSGs. WNL stars with very thin hydrogen envelopes may produce He-rich CSM with a surface helium abundance of $X_{\rm He}>0.6$. These models may therefore support lower-mass single-star progenitor candidates for SNe Ibn, representing the terminal explosions of H-poor WR stars embedded in He-rich CSM \citep{2007ApJ...657L.105F,2007Natur.447..829P,2022ApJ...927...25M}. \citet{2026arXiv260525823C} suggested that the progenitor of SN~2022pda, which shows hybrid properties between SNe Ibn and IIn, was a late-type WR star with hydrogen, namely a WNL star. As further support, the CSM velocity measured for SN~2022pda, $1900\,{\rm km\,s^{-1}}$, is broadly consistent with the wind velocity inferred for WR~120, $1225\,{\rm km\,s^{-1}}$.

Although the two selected solutions correspond to different evolutionary timings, both imply that WR\,120 has already lost most of its original H-rich envelope before reaching the observed HR diagram position. The PoWR analysis gives a H-free surface composition for WR~120 and classifies it as a WNE-s object \citep{Hamann2019}. In contrast, the two representative evolutionary models still retain relatively high surface H mass fractions, with $X_{\rm H}\simeq 0.4$, but only a very thin H-rich envelope remains. This suggests that a large fraction of the H-rich envelope was removed during the previous evolution and may have contributed to the formation of H-rich circumstellar material. In this interpretation, the absence of prominent hydrogen emission lines does not necessarily require a completely H-free evolutionary structure, but may instead reflect the small remaining envelope mass and the advanced stripping history. The low surface rotational velocities predicted by the models are also consistent with the narrow emission-line characteristics of WNE-s stars. Therefore, despite the atmosphere-based H-free classification, WR~120 can still be regarded as a late-type WN object that is evolutionarily connected to the WNL regime. More generally, our models suggest that H-free WN7--9 stars should be discussed together with WNL stars when the classification is based primarily on their late WN spectral subtype.


\begin{figure*}[ht!]
\centering
    \includegraphics[width=0.495\textwidth]{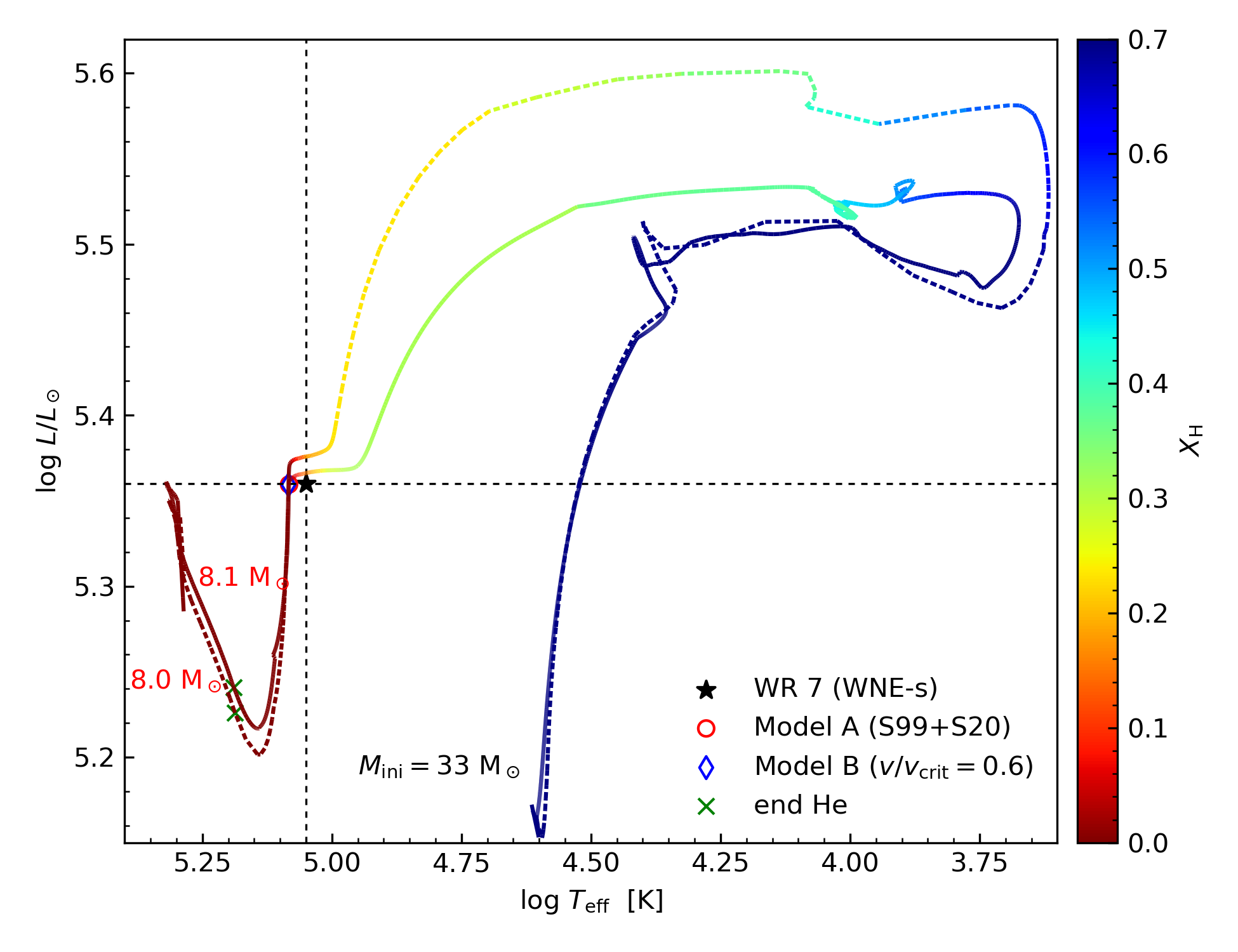}
    \includegraphics[width=0.495\textwidth]{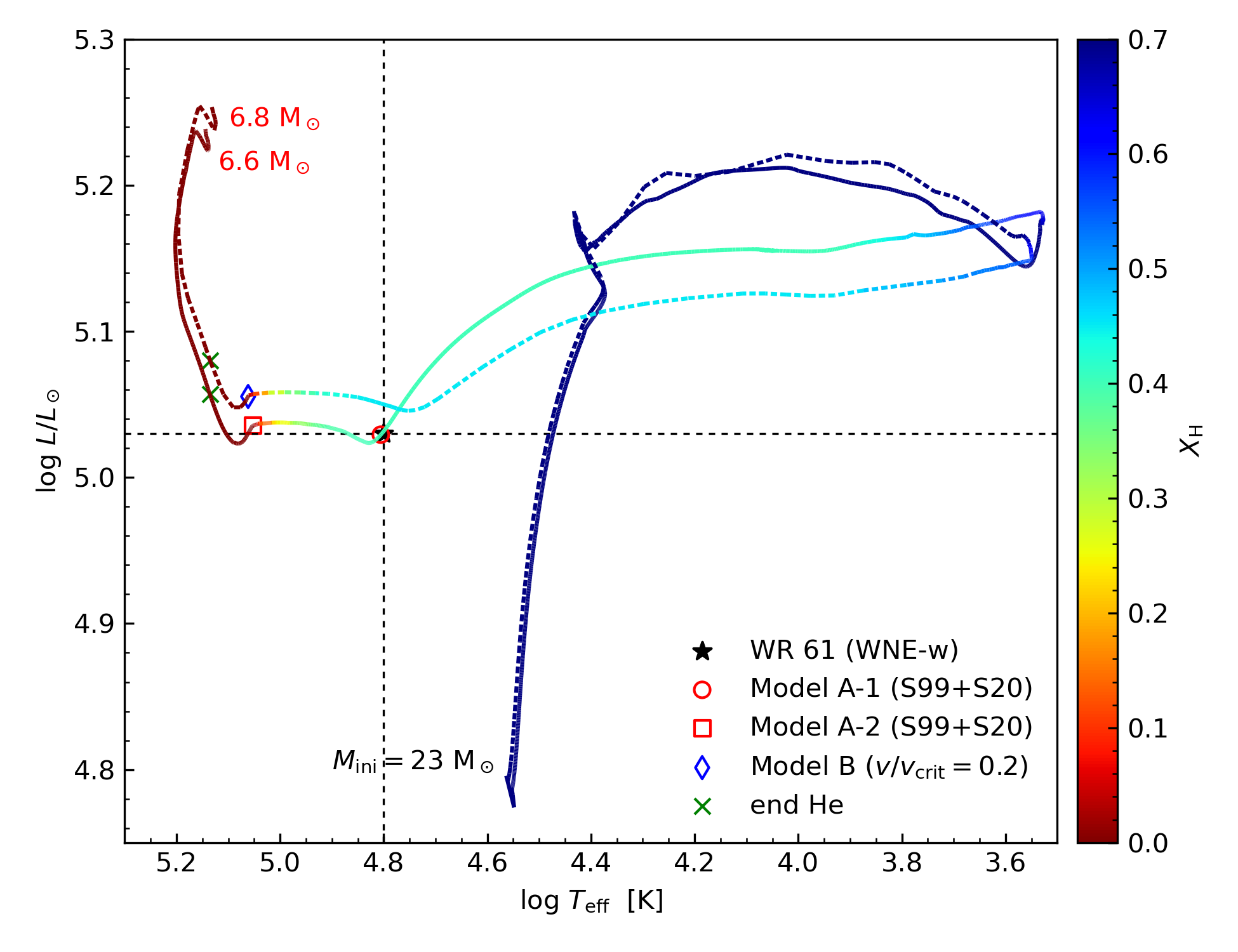}
\caption{
Same type of HR diagram comparison as in Fig.~\ref{WNL}, but for the two WNE objects, WR\,7 and WR\,61. 
The tracks are colour-coded by \(X_{\rm H}\), and the highlighted symbols mark the representative model points selected for comparison with the observed positions.
}
\label{WNE}
\end{figure*}

\subsubsection{WNE stars}

Normally, WNE stars are early-type WN (WN2--5) stars without hydrogen. Recent spectroscopic analyses have shown that several H-poor WN stars with late-type spectral morphology are more appropriately assigned to the WNE class based on their atmosphere solutions, particularly in the Large Magellanic Cloud (LMC) \citep{Hamann2019}. In addition, several H-free WN6 stars are commonly assigned to the WNE type.
The massive-star grids at $Z=0.014$ of \citet{Georgy2012} produce almost no WNE stars when the criterion $X_{\rm H}<10^{-5}$ is adopted. To better match the relatively high fraction of WNE stars in the Galactic WR catalogue \citep{2015wrs..conf...21C,2015MNRAS.447.2322R,Li2023}, we classify WN stars with $X_{\rm H}<0.1$ as WNE stars in this work. 

As WNE stars possess only very thin H-rich envelopes, their surface layers lie close to the helium core. 
Their surface rotation can therefore be strongly affected by angular-momentum redistribution and wind-driven angular-momentum loss. 
Spectroscopic analyses divide WNE stars into strong-lined (WNE-s) and weak-lined (WNE-w) objects, which are generally associated with rapidly and slowly rotating WNE stars, respectively. 
Recent calculations including internal gravity waves and the revised magnetic Tayler instability show that efficient angular-momentum transport can produce slowly rotating WNE stars \citep{2026ApJ...998...96S}. 
Thus, the WNE-w classification should not be interpreted only in terms of the initial rotation rate, but also in terms of internal angular-momentum transport and subsequent wind braking. 
The physical parameters of the selected evolutionary models are also presented in Table~\ref{tab:Table_WN}.

\vspace{0.2cm}
\noindent \textbf{WR~7} is characterized by a relatively high luminosity and effective temperature among the low-luminosity WR samples.
The selected model points with different initial rotational velocities lie very close to the observed HR diagram position and occur at very low surface H mass fraction.
This makes WR\,7 one of the clearest cases in the sample where the single-star tracks reach a plausible H-poor WNE phase at the observed luminosity and temperature.
The best-fitting Models~A and B for WR~7 listed in Table~\ref{tab:Table_WN} exhibit very similar properties.
Both models predict stellar radii and masses that are slightly lower than the observed values, while their mass-loss rates are in excellent agreement with the observations.
The rotational velocities predicted by these models exceed $200\,{\rm km\,s^{-1}}$, supporting the interpretation of WR~7 as a WNE-s star with strong emission-line features.

The difference between Models~A and B mainly reflects the effect of the higher initial rotational velocity adopted in Model~B.
During the RSG excursion, the faster rotating model reaches a lower minimum effective temperature, because enhanced rotational mixing favours stronger envelope expansion.
In the subsequent blueward evolution, this model evolves at slightly higher luminosity and removes its surface hydrogen more rapidly.
This behaviour is expected because stronger internal rotational mixing builds a more massive central helium core.
By the time the model enters the WNE phase, the residual H-rich envelope has been almost completely stripped and the He-rich core is exposed.
As a result, Model~B retains a higher equatorial surface velocity than Model~A at the selected WNE point.

\vspace{0.2cm}
\noindent \textbf{WR~61} is more difficult to reproduce with a single evolutionary point.
Model~A-1 provides the closest match to the observed HR diagram position, with a luminosity and temperature close to the atmosphere-derived values of WR\,61.
However, its surface H mass fraction remains high, $X_{\rm H}\simeq 0.40$, and it still retains a substantial H-rich envelope.
Together with its very low equatorial surface velocity, this indicates that Model~A-1 is still in an early WNL-like phase rather than in a genuine H-free WNE phase.
Although its slow surface rotation is qualitatively consistent with the weak-lined WN5-w spectrum of WR\,61, its surface composition makes it an unreliable match to the adopted H-free WNE classification.

Models~A-2 and B provide a different compromise.
They are much more consistent with the required WNE-like surface composition, with very low or vanishing surface H mass fraction and almost no remaining H-rich envelope.
However, both selected points are displaced toward much hotter effective temperatures, lying at $T\simeq 113$--$115\,{\rm kK}$ compared with the atmosphere-derived $T_\ast=63.1\,{\rm kK}$.
They also have substantially higher surface rotational velocities than Model~A-1, which is more naturally associated with strong-lined WNE behaviour than with the weak-lined WN5-w subtype of WR\,61.
Thus, WR\,61 illustrates a residual degeneracy between HR diagram agreement and subtype/composition agreement: Model~A-1 matches the HR diagram position better but remains WNL-like, whereas Models~A-2 and B are compositionally closer to the WNE phase but are too hot and less consistent with the weak-lined spectral morphology.

Part of this temperature tension may reflect the known difference between atmosphere-based and evolutionary temperature definitions.
The PoWR stellar temperature $T_\ast$ is defined at a Rosseland optical depth of 20 \citep{Hamann2006}, whereas the effective temperature variable in the evolutionary calculations refers to the photospheric optical depth.
For weak-lined WN stars such as WR\,61, this mapping carries additional uncertainty because the wind optical-depth structure is less well constrained than for stars with stronger emission.
A systematic offset between PoWR-derived $T_\ast$ values and evolutionary-track temperatures for WNE stars was also noted by \citet{Hamann2006}.
We therefore retain WR\,61 only as a marginal WNE comparison object, rather than as a secure single-star match.
Compared with the WNC stars discussed below, the WNE stars remain less restrictive because their main requirement is the exposure of a H-poor WN surface, rather than a short-lived mixed WNC composition.

Table~\ref{tab:Table_WN} lists the representative evolutionary model points selected for the WN stars, while the adopted atmosphere-derived parameters are given in Table~\ref{tab:Table_LowLWR}.
For most WN objects, Models~A and B are chosen to bracket the HR diagram-proximate solutions.
For WR\,61, we additionally distinguish Model~A-1 and Model~A-2 in order to separate the HR diagram-proximate WNL-like solution from the hotter but compositionally more WNE-like solution.
WR\,120 is treated separately because its PoWR atmosphere solution is H-free, whereas the selected evolutionary models still retain a very thin H-rich envelope; this makes it a useful test case for connecting H-free late-type WN spectra with WNL-like evolutionary states.
The differences among the selected model points primarily reflect the residual degeneracy in evolutionary timing, surface composition, and track family at fixed observed $(L,T_\ast)$, and they are carried forward as an estimate of model-selection uncertainty in the subsequent wind-scaling analysis.

Using the $(M,R)$ values from the representative model points in Table~\ref{tab:Table_WN}, we compute $v_{\rm esc}$ (and, when applicable, $v_{\rm esc,eff}$) and evaluate the wind-scaling diagnostics $v_\infty/v_{\rm esc}$, $\eta_{\rm wind}$, and $L_{\rm wind}$ following Section~\ref{subsec:vesc}. For observed stars we list the terminal wind speed $v_\infty$, whereas for evolutionary model points we report the corresponding equatorial surface rotation speed $v_{\rm eq}$ in the same column.
For the WN subsample, these diagnostics serve as an internal reference for the low-luminosity analysis: they quantify how sensitive the inferred wind ratios are to the remaining A/B timing degeneracy at fixed $(L,T_\ast)$, and they provide a baseline against which the low-luminosity WNC and WC objects can be assessed.
Any significant tensions between the observed wind parameters and the evolutionary solutions, as indicated by extreme values of $\eta_{\rm wind}$ or atypical $v_\infty/v_{\rm esc}$, will be highlighted alongside the corresponding stars in the subsequent subtype sections.


\begin{figure*}[ht!]
\centering
\includegraphics[width=0.495\textwidth]{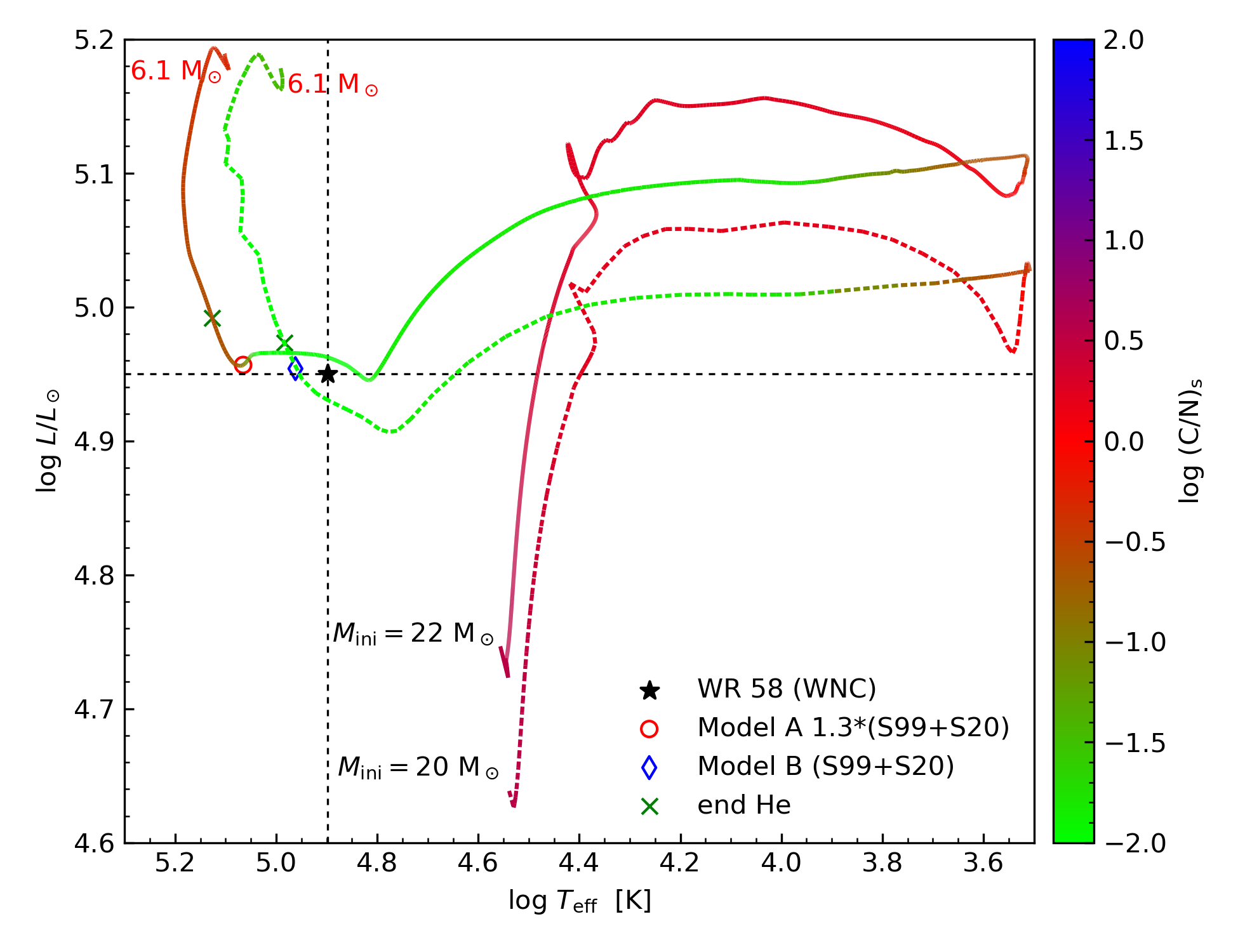}
\includegraphics[width=0.495\textwidth]{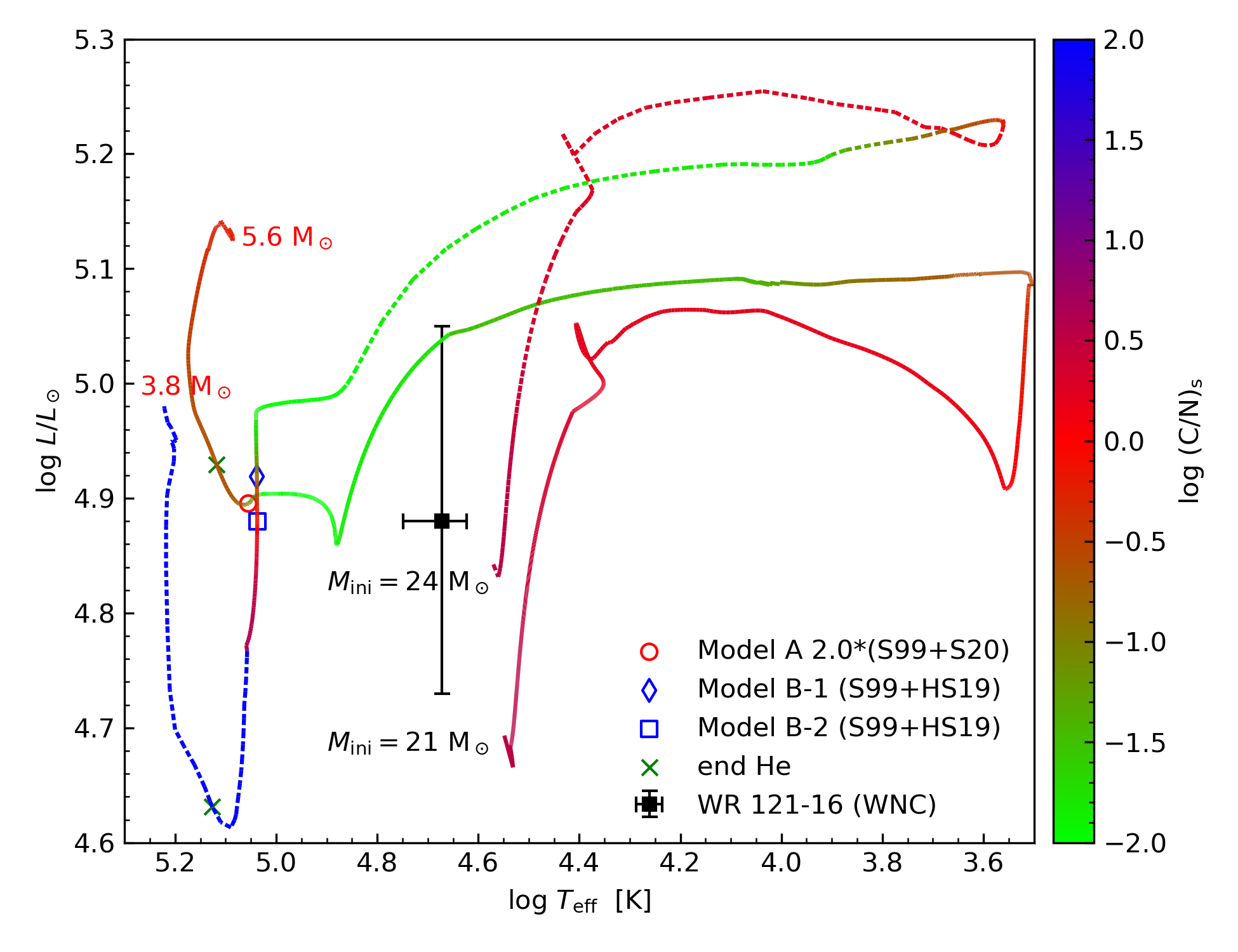}
\caption{
HR diagram comparisons for the two WNC objects, WR\,58 and WR\,121--16. 
The colour scale indicates the surface C/N mass ratio, expressed as \(\log(X_{\rm C}/X_{\rm N})\), along the tracks.
The highlighted symbols mark the representative WNC-like model points, while the green crosses indicate the end of core He-burning. The black star and the black square with error bars show the adopted observed positions.
}
\label{WNC}
\end{figure*}
   
\subsection{WNC stars}\label{subsec:results_wnc}

The WNC comparisons are shown in Figure~\ref{WNC}. 
WNC stars are defined here using the surface C/N mass ratio, \(0.1<X_{\rm C}/X_{\rm N}<10\) \citep{2013ApJ...764...21C,Peng2022LowLWC,Li2023}. The colours along the evolutionary tracks represent \(\log(X_{\rm C}/X_{\rm N})\). The red segments, with \(-1<\log(X_{\rm C}/X_{\rm N})<1\), mark the WNC phase that appears after the stars evolve blueward in the HR diagram. The physical parameters of the WNC models are presented in Table~\ref{tab:Table_WNC}.

Because WNC stars simultaneously exhibit the products of both hydrogen and helium burning, the mass range over which they occur provides a key diagnostic of the internal mixing processes in massive stars. The grids of WR stars with the $k-\omega$ model predict that WNC stars can form over an initial mass range of $15-35\, M_\odot$ \citep{2019ApJ...870...77L,Li2023}. This range can successfully account for the most massive WNC star in the sample of \citet{Sander2019}, WR,126, which has a current mass of $35.9\, M_\odot$. However, the predicted lower mass limit is obviously higher than the masses inferred for the low-luminosity WNC stars, which are only $7-9\, M_\odot$ (see Table \ref{tab:Table_LowLWR}, \citet{Sander2019,zhang2020}). Only 14 WNC stars are currently known in the Milky Way, and three of them belong to the low-luminosity WR stars. Given the rarity of these objects, constructing evolutionary models that can successfully reproduce such low-luminosity (and hence low-mass) WNC stars is crucial for constraining internal mixing processes and mass-loss histories, thereby helping to refine current models of massive star evolution.

\begin{table*}[t]
\centering
\caption{Observed parameters and representative Model~A and Model~B solutions for the selected low-luminosity WNC stars.}
\label{tab:Table_WNC}
\small
\setlength{\tabcolsep}{4pt}
\begin{tabular}{lllccccccccccccr}
\hline \hline
Model & Subtype & MLR & $\eta_{\rm Dutch}$ & log $L$ & $T$ & $R$ & $M$ & $M_{\rm ini}$ & $X_{\rm He}$ & $X_{\rm N}$ & $X_{\rm C}$ & $Y_{\rm c}$ & $v_{\rm eq}$ & log $\dot{M}$ & Age  \\
&  &  &  & ($L_\odot$) & (kK) & ($R_\odot$) & ($M_\odot$) & ($M_\odot$) & \multicolumn{4}{c}{(mass frac. \%)} & ($\rm km\ s^{-1}$) & ($M_\odot \ \rm yr^{-1}$) & (Myr) \\
(1) & (2) & (3) & (4) & (5) & (6) & (7) & (8) & (9) & (10) & (11) & (12) & (13) & (14) & (15) & (16) \\
\hline
WR~58  & WN4/WCE & & & 4.95 & 79.0 & 1.61 & 8.4 &  &  &  &  &  &  & -4.95 & \\
A & WNC & S99+S20 & 1.3 & 4.96 & 116.3 & 0.74 & 6.51 & 22 & 98 & 1.27 & 0.14 & 3 & 156 & -5.40 & 8.56 \\
B & WNL & S99+S20 & 1.0 & 4.95 & 91.6 & 1.19 & 6.33 & 20 & 59 & 1.07 & 0.01 & 0 & 145 & -5.45 & 9.48 \\
WR~121--16 & WN7o/WC & &  & 4.88 & 47 & 4.14 & 7.1 & & 98 & $1.5^{+1}_{-1}$ & $0.2^{+0.1}_{-0.1}$ &  &  & -4.97 &  \\
A & WNC & S99+S20 & 2.0 & 4.90 & 113.6 & 0.73 & 6.12 & 21 & 98 & 1.26 & 0.13 & 3 & 160 & -5.45 & 9.04 \\
B-1 & WNC & S99+HS19 & 1.0 & 4.92 & 109.2 & 0.81 & 6.90 & 24 & 98 & 1.24 & 0.20 & 35 & 125 & -5.08 & 7.55 \\
B-2 & WNC & S99+HS19 & 1.0 & 4.88 & 109.1 & 0.77 & 6.43 & 24 & 97 & 1.18 & 1.01 & 27 & 95 & -5.13 & 7.61 \\
\hline
\end{tabular}
\tablefoot{
The parameter $\eta_{\rm Dutch}$ is the overall scaling factor for the Dutch scheme in MESA. However, in this work the mass-loss prescriptions during the RSG and WR phases are replaced by the S99/Y23 and S20/HS19 prescriptions, respectively.
MLR denotes the adopted mass-loss-rate prescription.
The observed WNC parameters and their literature sources are given in Table~\ref{tab:Table_LowLWR}. Columns (10)--(12) list surface mass fractions in percent. All C/N ratios quoted in the text are mass ratios, defined as $X_{\rm C}/X_{\rm N}$.
}
\end{table*}

\vspace{0.2cm}
\noindent \textbf{WR\,58} can be approached by the selected tracks in the HR diagram, and the representative model points lie close to the observed luminosity. 
However, the WNC classification imposes a stronger condition than HR diagram proximity alone, because the model must expose material with both WN-like and WC-like abundance signatures at the correct time.

Model~B, with initial mass $20\,M_\odot$, matches the observed position at the end of core He-burning. However, it remains in the WNL phase ($X_{\rm H}\approx 0.4$), and with a surface N mass fraction approximately two orders of magnitude higher than the surface C mass fraction. 
As described in Sect.~\ref{subsec:global_hrd}, the luminosities predicted by the evolutionary models remain slightly higher than the observed values for WNC stars.
To reach a WNC solution at lower luminosity, we increased the overall scaling factor of $\eta_{\rm Dutch}$ to 1.30. The best-fitting model~A has $X_{\rm C}/X_{\rm N}\simeq0.11$ [$\log(X_{\rm C}/X_{\rm N})\simeq -0.96$] and a central He mass fraction of $Y_{\rm c}=0.03$. This indicates that WR~58 may have just entered the WNC phase. Meanwhile, the star is already approaching the end of core He burning. 

The best-fitting model yields a mass-loss rate of $\log \,\dot{M}=-5.40\, M_\odot \,\rm yr^{-1}$. Even after increasing the scaling factor by a factor of 1.3, the S20 prescription still significantly underestimates the $\dot{M}$ of low-luminosity WR stars. Furthermore, the discrepancy becomes increasingly pronounced toward lower luminosities.

\vspace{0.2cm}
\noindent \textbf{WR\,121--16} is more problematic. Despite being a recently discovered WNC star with detailed surface parameters provided by \citet{zhang2020}, no evolutionary model in our grid can simultaneously reproduce all of its observed properties.
Although the selected tracks reach a comparable luminosity, the representative WNC-like model points are substantially hotter and more compact than the observed position.
The observed star has \(T_\ast=47.0\,{\rm kK}\) and \(R=4.14\,R_\odot\), whereas both representative WNC solutions have effective temperatures of \(T_{\rm eff}\sim110\,{\rm kK}\) and radii below \(1\,R_\odot\).

Model~A, with an MLR of \(2.0\times({\rm S99+S20})\), reproduces the observed luminosity and reaches a WNC-like C/N mass ratio, \(X_{\rm C}/X_{\rm N}\simeq0.10\) [\(\log(X_{\rm C}/X_{\rm N})\simeq -0.99\)].
Its central He mass fraction, $Y_{\rm c}=0.03$, indicates that this solution is already close to the end of core He burning, similar to the preferred solution for WR\,58.
However, this model is still much hotter than WR\,121--16 and predicts a lower mass-loss rate than observed.

Model~B-1 and Model~B-2, with the HS19 wind prescription in the WR phase, give higher surface C mass fractions, with $X_{\rm C}=0.20\%$ and $X_{\rm C}=1.01\%$, respectively. Moreover, their predicted mass-loss rates are very close to the observed value.
Nevertheless, both solutions remain far too hot compared with the observed WNC position and have much smaller radii.
Therefore, changing the wind prescription can improve the surface-composition and wind-rate consistency, but it does not remove the main HR diagram discrepancy for WR\,121--16.

Models~B-1 and B-2 both have initial masses of $24\,M_\odot$ and are in the late stages of core He burning ($Y_{\rm c}<0.4$). Their ages are approximately 7.6 Myr, significantly younger than that of Model~A (9.04 Myr). Model~B-1 yields a surface C mass fraction consistent with the observations and a C/N mass ratio of $X_{\rm C}/X_{\rm N}\simeq0.16$. 
This indicates that WR,121--16 is likely at the onset of the WNC phase, having only recently begun to expose He-burning products and exhibit significant surface carbon enrichment. Model~B-2 has a lower luminosity and therefore provides a closer match to the observed luminosity of WR\,121--16. 
Its surface C mass fraction is approximately five times higher than the observed value, whereas the surface N mass fraction decreases only slightly. As a result, Model~B-2 reaches a C/N mass ratio of \(X_{\rm C}/X_{\rm N}\simeq0.86\), indicating that it has already evolved to an intermediate stage of the WNC phase.

This makes WR\,121--16 one of the clearest cases where the single-star tracks do not simultaneously reproduce the observed temperature, luminosity, radius, wind properties, and transitional surface composition.
The WNC stars therefore remain the strongest evidence in this sample that wind revision alone is insufficient, and that an additional stripping or mixing channel, such as binary mass transfer, may be required.

\subsection{WC stars}\label{subsec:results_wc}

\begin{figure}[ht!]
\centering
\includegraphics[width=\hsize]{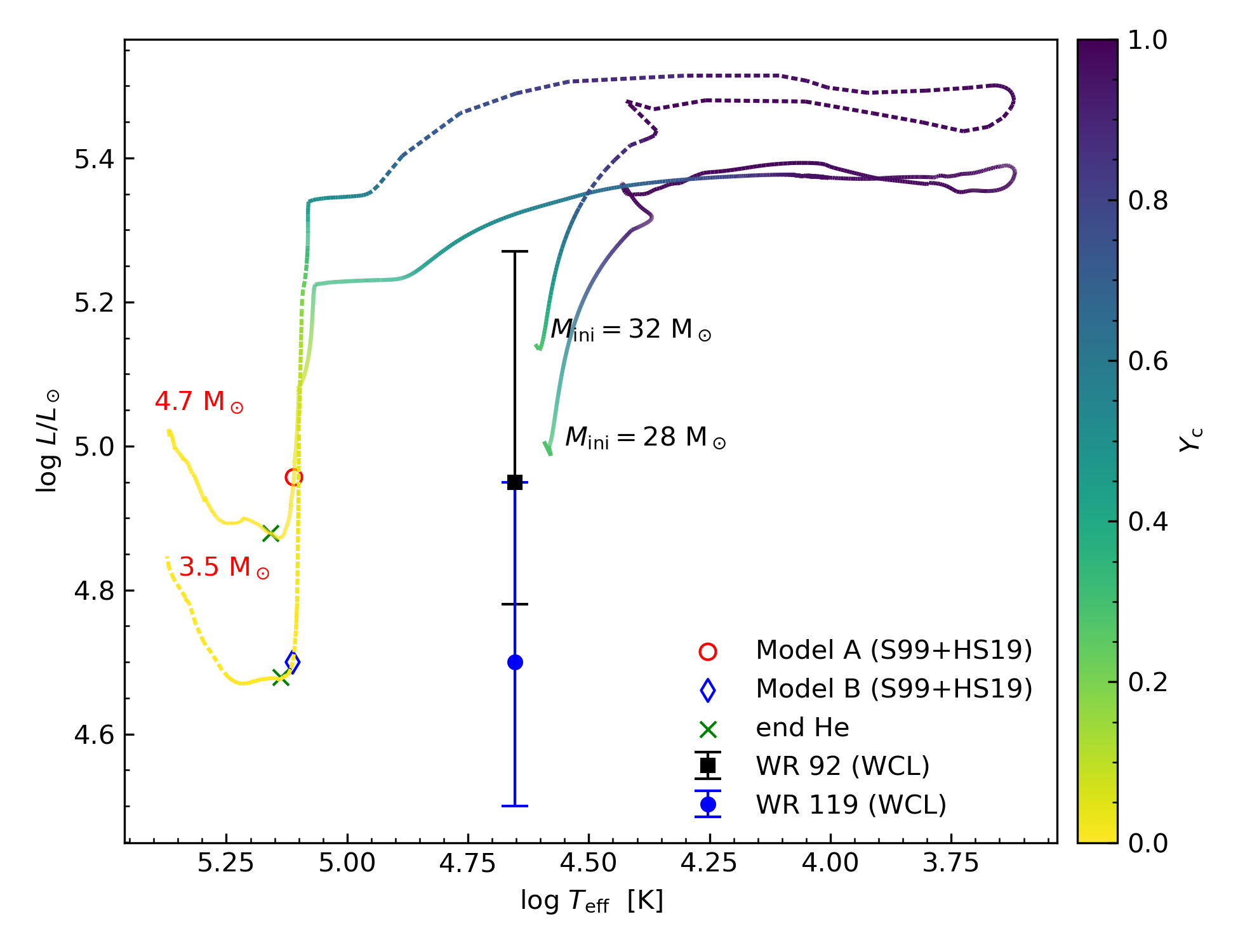}
\includegraphics[width=\hsize]{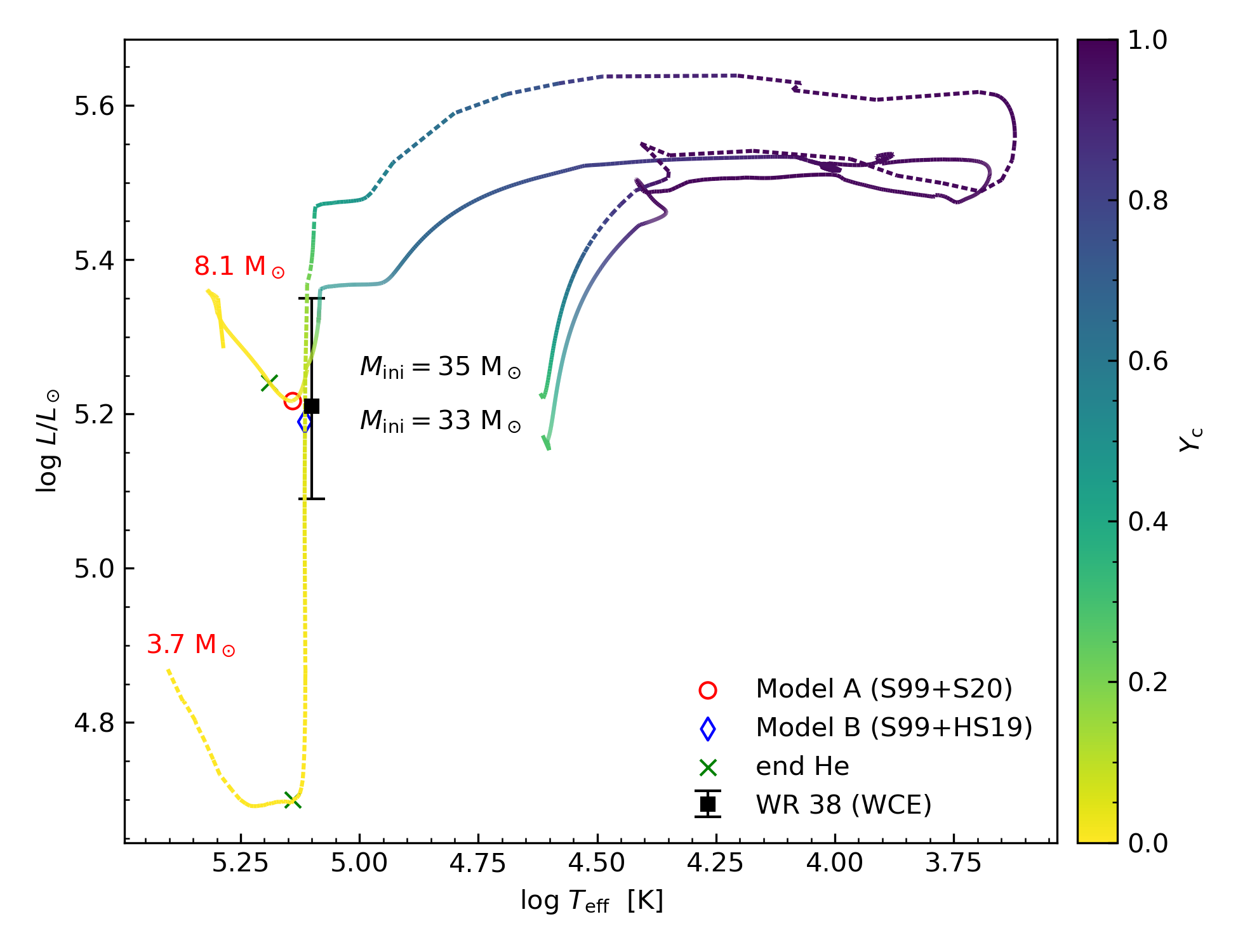}
\caption{
HR diagram comparisons for the WC objects in the updated sample. 
The upper panel shows the low-luminosity WCL stars WR\,92 and WR\,119 compared with HS19-based WC tracks, while the lower panel shows the WCE star WR\,38. 
Tracks are colour-coded by the central helium mass fraction, \(Y_{\rm c}\), and highlighted symbols mark the representative model points.
}
\label{WCL}
\end{figure}

The updated WC comparison is shown in Figure~\ref{WCL}. 
The upper panel focuses on the two low-luminosity WCL stars WR\,92 and WR\,119. 
With the HS19 WR-wind prescription, the tracks can reach the luminosity range of these objects, including the lower luminosity of WR\,119. 
Those models follow the massive star scenario described by \citet{1994A&AS..103...97M,Georgy2012,Peng2022LowLWC}, but originate from relatively lower initial masses ($M_{\rm ini}<40\,M_\odot$). 
Unlike the case of the WNC stars, for which the predicted mass-loss rates tend to be lower than the observed values, the HS19 prescription (see Eq.~\ref{eq:HS19_WC}) significantly overestimates the mass-loss rates of low-luminosity WC stars.
The selected model points remain significantly hotter than the observed WCL positions. The improvement should therefore be interpreted mainly as a luminosity-side improvement, not as a complete match to the cool late-type WC regime. The physical parameters of the WC models are presented in Table~\ref{tab:Table_WC}.

\begin{table*}[t]
\centering
\caption{ Observed parameters and representative Model~A and Model~B solutions for the selected low-luminosity WC stars. }
\label{tab:Table_WC}
\small
\setlength{\tabcolsep}{5pt}
\begin{tabular}{lllccccccccccccr}
\hline \hline
Model & Subtype & MLR & log $L$ & $T$ & $R$ & $M$ & $M_{\rm ini}$ & $X_{\rm C}$ & $X_{\rm O}$ & $X_{\rm He}$ & $Y_{\rm c}$ & $M_{\rm He}$ & $v_{\rm eq}$ & log $\dot{M}$ & Age \\
&  &  & ($L_\odot$) & (kK) & ($R_\odot$) & ($M_\odot$) & ($M_\odot$) & \multicolumn{4}{c}{(mass fraction, \%)} & ($M_\odot$) & ($\rm km\ s^{-1}$) & ($M_\odot \ \rm yr^{-1}$) & (Myr) \\
(1) & (2) & (3) & (4) & (5) & (6) & (7) & (8) & (9) & (10) & (11) & (12) & (13) & (14) & (15) & (16) \\
\hline
WR~92  & WC9 & & 4.95 & 45.0  & 4.97 & 7.70 &  &  &  &  &  &  &  & -5.0 & \\
A  & WC & S99+HS19 & 4.96 & 128.8 & 0.60 & 6.29 & 28 & 52 & 27 & 18  & 2 & 0.26 & 90 & -4.59 & 6.82 \\
WR~119 & WC9d & & 4.70 & 45.0  & 3.70 & 5.8 &  &  &  &  &  &  &  & -5.13 & \\
B  & WC & S99+HS19 & 4.70 & 129.6 & 0.44 & 4.64 & 32 & 41 & 51 & 6 & 0 & 0.06 & 65 & -4.40 & 6.18 \\
\hline
WR~38  & WC4  & & 5.21 & 126.0 & 0.85 & 10.4 &  &  &  &  &  &  &  & -4.66 & \\
A  & WC & S99+S20  & 5.22 & 138.1 & 0.71 & 8.48 & 33 & 50 & 15 & 33 & 1 & 0.41 & 84 & -4.92 & 6.00 \\
B  & WC & S99+HS19 & 5.19 & 130.2 & 0.77 & 8.48 & 35 & 51 & 28 & 19 & 6 & 0.68 & 82 & -4.44 & 5.71 \\
\hline
\end{tabular}
\tablefoot{
The surface mass fractions are listed only for the evolutionary model points. 
MLR denotes the adopted mass-loss-rate prescription.
The observed WC parameters and their literature sources are given in Table~\ref{tab:Table_LowLWR}.
Col. (9)--(11) list the surface mass fractions of C, O, and He, respectively, expressed in percent.
Col. (12) gives the central He mass fraction, and Col. (13) lists the total mass of He.
}
\end{table*}

\vspace{0.2cm}
\noindent \textbf{WR~119} is classified as a single WCL (WC9d) star according to the WR catalogue. It is the least luminous WC sample \citet{Sander2019}, with $\log L/L_\odot=4.7$ and a current mass of $5.8\,M_\odot$. 
Model~B, with a slightly higher initial mass of $32\,M_\odot$, is able to reach the region occupied by the least luminous WC stars. Its larger convective core during core He burning allows it to enter the WC phase earlier, leading to enhanced mass loss over a longer period and, consequently, to the formation of a lower-luminosity WC star.

However, the WC stars produced through this single-star channel are characterized by extremely low masses ($<5.0\,M_\odot$) and very compact radii ($\sim 0.44\,R_\odot$), which are substantially lower than those inferred from observations. This indicates that WR~119 most likely formed through a close-binary channel. Moreover, \citet{2020A&A...641A..26D} found a value of $\Delta\,{\rm RV}=11.0\,{\rm km\,s^{-1}}$ for WR~119 and classified it as a binary.

\vspace{0.2cm}
\noindent \textbf{WR~38} is a single WCE (WC4) object and shown in the lower panel of Figure~\ref{WCL}. 
In contrast to WR\,92 and WR\,119, WR\,38 is much better reproduced in the HR diagram: both the luminosity and effective temperature of the representative model points are close to the observed position. 
This indicates that the WCE case is not primarily an HR diagram problem in the present comparison.
Both Model~A and Model~B, computed with the S20/HS19 prescription during the WR phase, reproduce the observed properties reasonably well. Their logarithmic mass-loss rates are close to the observed value, \(\log(\dot{M}/M_\odot\,{\rm yr}^{-1})=-4.66\), and their radii are only slightly smaller than the observed value of \(0.85\,R_\odot\).

The low-mass WC stars ($<10\,M_\odot$) produced by our models may be relevant to the low-luminosity progenitor scenario proposed for some Type Ibc supernovae, although their radii and effective temperatures remain difficult to reconcile with the observed properties of low-luminosity WCL stars \citep{2009ARA&A..47...63S,Sander2012,Georgy2012}. The final masses of these WC models are extremely low ($<5\,M_\odot$), while the helium mass in the ejecta may be lower than the values listed in Table~\ref{tab:Table_WC} ($<0.6\,M_\odot$). Consequently, a large fraction of these models are expected to end their evolution as low-mass, low-luminosity, naked carbon--oxygen (CO) core stars and eventually explode as Type Ic supernovae.

The updated WC result is therefore mixed. 
The HS19 prescription helps the models reach the faint WCL luminosity range, but the cool temperatures of WR\,92 and WR\,119 remain difficult to reproduce. 
WR\,38, on the other hand, is compatible with the selected WC-stage tracks in the HR diagram. 
Thus, the WC discussion should distinguish between the low-luminosity WCL problem and the WCE case, rather than treating all WC stars as having the same level of tension.



\subsection{Wind-budget diagnostic for the nine-source sample}
\label{subsec:wind_budget_diagnostic}

\begin{figure*}[htbp]
\centering
\includegraphics[width=0.88\textwidth]{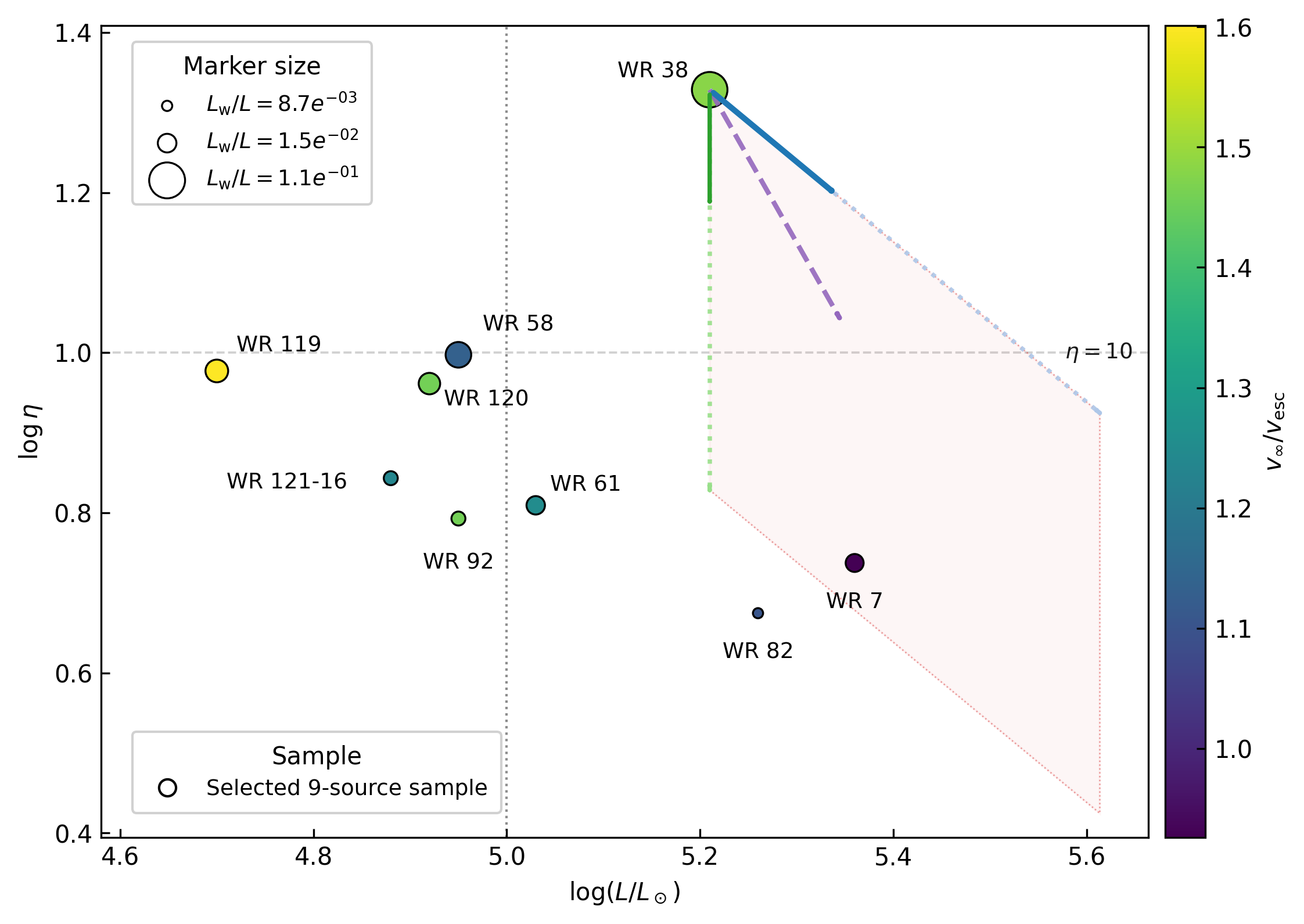}
\caption{
Wind-budget diagnostic plane for the nine-source sample. 
The vertical axis gives the wind-efficiency parameter \(\eta_{\rm wind}=\dot{M}v_\infty/(L/c)\), while the horizontal axis gives the stellar luminosity. 
Symbol size traces the mechanical wind luminosity relative to the stellar luminosity, \(L_{\rm wind}/L\), and colour indicates \(v_\infty/v_{\rm esc}\). 
The horizontal dashed line marks \(\eta_{\rm wind}=10\), used here as a heuristic indicator of a high wind-momentum requirement rather than as a sharp physical boundary, and the vertical dotted line marks \(\log(L/L_\odot)=5.0\). 
For WR\,38, arrows illustrate the effect of distance/extinction and relative clumping corrections on the inferred \((\log L,\log\eta_{\rm wind})\) position. 
The clumping corrections are shown relative to the atmosphere-analysis value \(f_{\rm cl,assumed}=0.1\). 
This diagnostic highlights that WR\,38 is not mainly an HR diagram outlier, but remains the most demanding object in terms of wind momentum.
}
\label{fig:wind_budget_systematics}
\end{figure*}

To complement the HR diagram comparison, we place the nine stars in a common wind-budget diagnostic plane in Fig.~\ref{fig:wind_budget_systematics}. 
This comparison is not used as an additional fitting constraint, but as a consistency check on whether the selected evolutionary solutions can also accommodate the observed wind momentum and mechanical wind power. 
The WN reference objects occupy the lower-\(\eta_{\rm wind}\) part of the diagram, whereas the WNC stars lie close to the \(\eta_{\rm wind}\simeq10\) boundary, where multiple scattering and wind-density systematics become especially relevant for interpreting the momentum budget \citep{LamersCassinelli1999,Crowther2007,Puls2008}. 
Among the WC objects, WR\,92 and WR\,119 are not the most extreme cases in wind efficiency, despite their low luminosities and late WC subtypes. 
By contrast, WR\,38 stands out with the largest wind momentum requirement, \(\eta_{\rm wind}\simeq21\), even though it is the WC object best matched in the HR diagram.

The systematic shifts shown for WR\,38 illustrate why the wind-budget result should be interpreted as a diagnostic rather than as a sharp classification boundary. 
Increasing the adopted distance moves the object to higher luminosity and lower \(\eta_{\rm wind}\), while reducing the true clumping factor below the value assumed in the atmosphere analysis also lowers the inferred wind efficiency. 
Even so, WR\,38 remains the clearest case in the present sample where a good HR diagram match must still be checked against the wind-density requirement. 
This supports the conclusion that the WC comparison is not simply a luminosity problem, but a joint constraint involving luminosity, temperature, surface composition, and WR-like wind strength.

\begin{table}[htbp]
\centering
\caption{
Derived wind-budget diagnostics for the updated nine-object low-luminosity WR sample.
The input stellar and wind parameters are taken from Table~\ref{tab:Table_LowLWR}.
}
\label{tab:wnc_wc_winddiag}
\small
\setlength{\tabcolsep}{5pt}
\begin{tabular}{llcccc}
\hline \hline
WR & Subtype & $v_{\rm esc}$ & $v_\infty/v_{\rm esc}$ & $\eta_{\rm wind}$ & $L_{\rm wind}/L$ \\
 & & (km\,s$^{-1}$) & & & \\
\hline
82      & WN7(h)    & 995  & 1.11 & 4.73  & $8.7\times10^{-3}$ \\
120     & WN7 & 841  & 1.46 & 9.15  & $1.9\times10^{-2}$ \\
7       & WN4-s     & 1984 & 0.81 & 5.46  & $1.5\times10^{-2}$ \\
61      & WN5-w     & 1117 & 1.25 & 6.45  & $1.5\times10^{-2}$ \\
58      & WN4/WCE   & 1411 & 1.13 & 9.94  & $2.7\times10^{-2}$ \\
121--16 & WN7o/WC   & 809  & 1.24 & 6.97  & $1.2\times10^{-2}$ \\
38      & WC4       & 2161 & 1.48 & 21.30 & $1.1\times10^{-1}$ \\
92      & WC9       & 769  & 1.46 & 6.21  & $1.2\times10^{-2}$ \\
119     & WC9d      & 773  & 1.68 & 9.49  & $2.1\times10^{-2}$ \\
\hline
\end{tabular}
\tablefoot{
The escape speed is the classical escape speed computed from the adopted mass and radius. 
The diagnostics are used as consistency checks and are not additional fitting constraints.
}
\end{table}

\section{Discussion}
\label{sec:discussion}

\subsection{The WN stars as a reference scale}
\label{subsec:disc_wn}

The WN stars provide a useful reference for the comparison because they are the part of the sample for which the evolutionary tracks perform best.
For WR\,82, WR\,120, WR\,7, and WR\,61, the selected tracks reproduce, or at least approach, the observed HR diagram positions without the stronger wind-budget tensions found for the low-luminosity WNC and WC stars 
(Sects.~\ref{subsec:results_wnc}--\ref{subsec:wind_budget_diagnostic}). 
The main difficulty is therefore not the use of the comparison method itself, but whether the models can also reproduce the radius, surface composition, and wind properties of each subtype. 
The observed WN parameters require a consistent interpretation of spectral subtype, stellar temperature, radius, and surface composition \citep{Crowther2007,Hamann2019}. 
The evolutionary models must then reach the corresponding surface abundances and wind properties at modest luminosity \citep{Ekstrom2012,Georgy2012,Li2023}.

This distinction is already visible within the WN subsample.
WR\,82 is the cleanest H-rich WN reference case in the present sample. 
Its HR diagram match lies within a narrow range of initial masses, and the selected model radii are close to the observed value. 
However, the model surface hydrogen fractions remain higher than the adopted observed value. 
Thus, WR\,82 supports the leading-order single-star interpretation in luminosity, temperature, and radius, but also shows that the degree of envelope stripping is not reproduced exactly.

WR\,120 requires a more nuanced interpretation. Although its spectral subtype is WN7, the PoWR analysis gives no detectable surface hydrogen and assigns it to a WNE-w atmosphere solution \citep{Hamann2019}. However, this atmosphere-based classification does not necessarily imply that the corresponding evolutionary model must be completely H-free. The selected evolutionary points approach the observed HR diagram position while retaining a substantial surface hydrogen fraction, but the remaining H-rich envelope is already very thin. This suggests that WR\,120 may represent an advanced WNL-like evolutionary state in which most of the original H-rich envelope has been removed, leaving little hydrogen available to produce prominent emission features. In this sense, WR\,120 supports the idea that some H-free late-type WN7--9 spectra can still be evolutionarily connected to WNL stars, rather than being interpreted only as classical H-free WNE objects.

The WNE stars WR\,7 and WR\,61 further show that the H-poor WN regime is not homogeneous. 
WR\,7 provides the most successful H-poor WNE comparison, because the selected models have low surface H mass fraction and radii close to the observed value. 
WR\,61, however, illustrates a stronger degeneracy. 
The H-rich model point gives a radius closer to the observed value, whereas the H-poor model points better reproduce the depleted surface H mass fraction but are much more compact and hotter. 
This behaviour suggests that, for some weak-lined WN stars, the model may reproduce either the radius/temperature scale or the surface hydrogen depletion, but not always both simultaneously. 
WR\,61 also illustrates the practical limit of the temperature comparison for weak-lined WN stars: part of the large \(T_\ast\) offset may be related to the atmosphere-to-evolution temperature mapping \citep{Hamann2006}. 
The slow-rotation aspect of the WNE-w interpretation should be treated as an angular-momentum transport problem, not only as a surface-composition issue. 

The present evolutionary grid does not include the additional internal angular-momentum transport mechanisms explored by \citet{2026ApJ...998...96S}, such as internal gravity waves and the revised magnetic Tayler instability. 
These processes can efficiently redistribute angular momentum and produce slowly rotating WNE stars. 
The weak-lined morphology of WR\,61 may therefore be easier to reconcile with a H-poor WNE evolutionary state if such angular-momentum redistribution is included, although the HR diagram and radius discrepancies remain in the present comparison. 

Among the four WN objects, WR\,82 and WR\,7 provide the closest matches across luminosity, temperature, and surface composition, while WR\,120 and WR\,61 each preserve a residual tension---between the atmosphere-based H-free classification and the retained model envelope in the former, and between the HR diagram-proximate but WNL-like solution and the too-hot WNE-like solution in the latter.


\subsection{WNC stars as tests of the transition phase}
\label{subsec:disc_wnc}

The two WNC stars, WR\,58 and WR\,121--16, probe the critical transition between the WN and WC phases and therefore constitute sensitive tests of whether single-star models can expose the relevant chemical transition layer at the correct evolutionary time. 
Both objects can be placed on relatively low-\(M_{\rm ini}\) tracks in our grid 
(Sect.~\ref{subsec:results_wnc}; Table~\ref{tab:Table_WNC}), so their HR diagram positions alone do not rule out a single-star origin. 
This placement is important because WNC stars occupy a short-lived transition between the WN and WC sequences \citep{1991A&A...248..531L,Crowther2007,Sander2019}. 
They are also difficult to reproduce with standard single-star evolution at near-solar metallicity, especially in the low-luminosity regime \citep{zhang2020}.

The difficulty appears when the HR diagram match is compared with the radius and surface-composition constraints.
A successful model must not only reach the observed luminosity, but must also expose a WNC transition composition, remain consistent with the adopted H-free surface, and reproduce the observed radius or temperature scale. 
The comparison in Table~\ref{tab:Table_WNC} shows that these requirements are not met simultaneously.
For WR\,58, the model point with the more convincing WNC-like composition is substantially hotter and more compact than the atmosphere solution. 
Model~A has \(T\simeq116\,{\rm kK}\) and \(R=0.74\,R_\odot\), compared with the observed \(T_\ast=79\,{\rm kK}\) and \(R=1.61\,R_\odot\). 
Model~B gives a larger radius, \(R=1.19\,R_\odot\), but it is less convincing as a WNC-composition match. 
Thus, the apparent HR diagram agreement does not translate into a complete physical match.

The discrepancy is even clearer for WR\,121--16. 
The observed star is relatively cool and extended for a WNC object, with \(T_\ast=47\,{\rm kK}\) and \(R=4.14\,R_\odot\). 
By contrast, the selected model points reach a similar luminosity in much more compact and hotter configurations, with \(R\simeq0.73\)--\(0.81\,R_\odot\) and \(T\simeq109\)--114\,kK. 
At nearly fixed luminosity, this is not a minor offset: the small model radii directly imply temperatures that are too high compared with the atmosphere solution. 
The problem is therefore not simply that the tracks fail to reach the luminosity of WR\,121--16; rather, they reach that luminosity with the wrong radius and temperature scale.

This radius--composition tension is physically important. 
The WNC phase depends on the detailed structure of the chemical transition region outside the He-burning core. 
This makes it sensitive to the adopted treatment of convective-boundary mixing \citep{Herwig2000}. 
It is also affected by rotation and by the mass-loss history along the evolutionary track \citep{Ekstrom2012,Georgy2012,Li2023}. 
The present comparison suggests that the models can expose WNC-like surface abundances, but they do so in configurations that are too compact for the observed WNC atmosphere parameters, especially for WR\,121--16.

The WNC stars are thus best viewed as boundary cases rather than as ordinary extensions of the WN or WC sequences. 
They remain compatible with the single-star framework only if the model simultaneously exposes the WNC transition layer, reaches the observed HR diagram position, reproduces the radius or temperature scale, and sustains a WR-like wind at modest luminosity. 
This combination is not achieved robustly in the present grid. 
The WNC objects therefore suggest that changing the WR wind prescription alone may not be sufficient.
Additional mixing could help modify the surface abundance pattern near the WNC transition \citep{Herwig2000,Li2023}. 
Alternatively, binary mass exchange or post-interaction stripping may expose chemically evolved layers at lower luminosity \citep{Eldridge2017,Gotberg2018,Shenar2020}. 
Such channels, however, must still explain the relatively large observed radii and cool atmosphere temperatures of the WNC stars, especially WR\,121--16.



\subsection{WC stars: separating the WCL and WCE constraints}
\label{subsec:disc_wc}

The updated WC comparison shows that the WC part of the sample should not be treated as a single homogeneous problem. 
For the two low-luminosity WCL stars, WR\,92 and WR\,119, the HS19-based tracks alleviate the luminosity-side tension and can reach the observed faint WC luminosity range, including the very low luminosity of WR\,119. 
However, the selected model points remain substantially hotter than the observed WCL positions. 
The WCL comparison therefore suggests that the HS19 WR-wind prescription can help explain the luminosities of the faintest WC stars, consistent with the sensitivity of WR winds to luminosity and composition, but it does not by itself provide a complete match to their cool late-type WC temperatures \citep{Sander2020,Peng2022LowLWC}.

WR\,38 behaves differently. 
Its observed position is close to the selected WCE model points in the HR diagram, so the main difficulty is not the location in the \((L,T_\ast)\) plane. 
Instead, the wind-budget diagnostic in Fig.~\ref{fig:wind_budget_systematics} shows that WR\,38 is the most demanding object in terms of wind efficiency, with \(\eta_{\rm wind}\) well above the reference \(\eta_{\rm wind}=10\) level before systematic corrections. 
Thus, WR\,38 is most useful as a check on whether an HR diagram match can also satisfy the surface-composition and wind-density requirements of a WC star.

The WC comparison therefore points to different limitations for the WCL and WCE stars. 
The WCL stars show that the HS19 prescription helps populate the low-luminosity WC regime, but their cool observed temperatures remain a challenge. 
The WCE star WR\,38 is much better matched in the HR diagram, but still needs to satisfy the wind and composition diagnostics. 
The WC problem is therefore not simply a question of reaching low luminosity; it is a joint requirement involving luminosity, temperature, surface composition, and WR-like wind density.

\subsection{Implications for mixing, mass loss, and binary channels}
\label{subsec:disc_implications}

The subtype-level results above indicate that the limitations of the current grid are not set by a single observable, but by the simultaneous requirements on luminosity, temperature or radius, surface composition, and wind density.
The WN stars define the relatively well-behaved reference part of the sample, whereas the WNC and low-luminosity WC stars provide the most restrictive tests of the adopted mixing and wind prescriptions.

From the standpoint of stellar-evolution physics, two linked requirements emerge. 
First, the mixing prescription must create transition layers of the right extent and expose them at the right time to populate the WNC regime \citep{Herwig2000,Ekstrom2012,Georgy2012,Li2023}. 
Second, the adopted WR-phase mass-loss history must be strong enough to uncover WC-like surfaces at low luminosity, while still producing stars with temperatures and wind properties compatible with the observed subtype \citep{NugisLamers2000,Vink2001,Puls2008}. 
The updated WCL comparison suggests that HS19-like WR winds help with the luminosity requirement, but the remaining temperature offset shows that this is not yet a complete solution. This point is also illustrated by earlier high-mass evolutionary calculations, where a \(60\,M_\odot\) O--LBV--WR sequence could be stripped to a final mass of only a few solar masses, while a \(35\,M_\odot\) O--RSG--WR sequence retained a larger final mass \citep{1996A&A...316..133G,1996A&A...305..229G}. 
Such examples show that the mapping between present-day WR luminosity and \(M_{\rm ZAMS}\) is strongly model dependent, especially when enhanced LBV/WR mass loss is adopted.

Binary stripping provides a natural channel to populate the low-luminosity WNC and WC regime, since it can expose chemically evolved helium-star layers at luminosities below those predicted by standard single-star tracks.
Recent population-synthesis work further suggests that merger products may contribute to the faint WR population \citep{Li2024Merger}.
The present comparison does not require all low-luminosity WNC and WC stars to have a binary origin, but the simultaneous failure of luminosity, temperature, and wind-budget matching strengthens the case for invoking such channels in at least some objects.

At the same time, a binary-origin helium star should not automatically be equated with a classical WR star.
Recent theoretical and observational work shows that stripped-envelope stars span a wide sequence, from subdwarf-like objects to stars with genuinely WR-like spectra, and that the latter likely correspond to the subset with sufficiently dense and optically thick winds \citep{Gotberg2018,Drout2023,Yungelson2024,2025A&A...703A.243B}.
This caveat is important for the present problem.
A helium-star merger or other post-interaction remnant may be a viable route to a faint WC star, but not every such remnant is expected to display the strong WR-type emission needed for a formal WC classification \citep{Shenar2023Magnetar,Shenar2024}.

The staged comparison is therefore most useful not as a binary classifier, but as a way to identify where additional observational or theoretical work is most needed: the WN stars largely validate the method, the WNC objects expose its limits under the current grid, and the WCL/WCE distinction shows that the faint-WR problem is not monolithic.
Lower-luminosity WNC/WC stars are therefore particularly valuable benchmarks for future work on WR mixing and WR-phase mass loss \citep{Crowther2007,Li2023}. 
They are also useful for testing the contribution of binary products to the Galactic WR population \citep{Eldridge2017,Gotberg2018,Drout2023,Shenar2024}; all subtype-level conclusions drawn here rest on nine objects and will require revision as homogeneous atmosphere analyses extend to larger and fainter WR samples.


\subsection{Kinematic context from Gaia-based runaway screening}
\label{sec:kinematic_context}

\begin{table*}[htbp]
\centering
\small
\caption{
Derived Gaia-based kinematic runaway diagnostics for the WR subset with updated astrometric solutions.
}
\label{tab:wr_runaway_screen}
\resizebox{\textwidth}{!}{%
\begin{tabular}{llcccccl}
\hline \hline
Object & Subtype & $d_{50}$ & $V_{{\rm t,pec},50}$ 
& $P(V_{\rm t,pec}>30)$ & $P(|\eta_{\rm z}|>1)$ 
& Arthur quality & Kinematic classification \\
 & & (kpc) & (km\,s$^{-1}$) & & & & \\
\hline
WR~82      & WN7(h)      & 3.89 & 29.82 & 0.483 & 1.000 & pass & runaway by height \\
WR~120     & WN7/WNE-w   & 3.35 & 7.64  & 0.161 & 0.000 & fail & inconclusive (RUWE above cut) \\
WR~7       & WN4-s       & 4.28 & 54.27 & 1.000 & 0.000 & pass & runaway by velocity \\
WR~61      & WN5-w       & 6.72 & 67.55 & 1.000 & 1.000 & pass & runaway by velocity and height \\
WR~58      & WN4/WCE     & 6.95 & 32.33 & 0.748 & 1.000 & pass & runaway by height \\
WR~121--16 & WN7o/WC     & 8.42 & 40.25 & 1.000 & 1.000 & pass & runaway by velocity and height \\
WR~38      & WC4         & 7.06 & 8.36  & 0.034 & 0.025 & pass & not runaway in this screen \\
WR~92      & WC9         & 4.59 & 96.19 & 1.000 & 1.000 & pass & runaway by velocity and height \\
WR~119     & WC9d        & 4.31 & 56.15 & 1.000 & 0.949 & pass & runaway by velocity and height \\
\hline
\end{tabular}%
}
\tablefoot{
The screening follows the basic logic of \citet{Arthur2025}, but is applied here as a homogeneous supplementary test for the present nine-object sample rather than as a full reproduction of the Arthur et al. analysis.
The distance posterior adopts a direction-dependent exponentially decreasing space-density prior following \citet{BailerJones2018}.
``Arthur quality'' indicates whether the source passes the adopted astrometric cuts, where RUWE denotes the renormalized unit weight error: \(\mathrm{RUWE}<1.5\) and \(\varpi_{\rm corr}/\sigma_\varpi>5\).
}
\end{table*}

As a supplementary check on the evolutionary interpretation, we performed a homogeneous Gaia-based kinematic screen for the subset of stars for which the updated astrometric screening was available.
The purpose is not to redefine the staged model--data comparison, but to test whether the stars that appear most relevant in the HR diagram and wind-budget analysis are also distinguished kinematically. 
We adopt an Arthur-style screening procedure, in the sense that candidate runaways are identified through either a peculiar tangential velocity above \(30\,{\rm km\,s^{-1}}\) or a vertical displacement exceeding the local scale height of O- and B-type (OB) stars. 
The calculation uses Gaia Data Release 3 (DR3) astrometry \citep{GaiaCollaboration2023}, the Gaia parallax zero-point correction \citep{Lindegren2021}, a direction-dependent exponentially decreasing space-density distance prior following \citet{BailerJones2018}, \texttt{emcee} posterior sampling \citep{ForemanMackey2013}, and a warp/flare-based \(\eta_{\rm z}\) diagnostic following \citet{Li2019}. 
Gaia queries and coordinate handling were performed with \texttt{astroquery} and \texttt{astropy} \citep{Ginsburg2019,AstropyCollaboration2022}. 
Because our analysis is based primarily on tangential motions and height diagnostics, the resulting classifications should be viewed as a kinematic characterization of the sample rather than a substitute for full orbital reconstructions.
The results for the selected WR sample are listed in Table~\ref{tab:wr_runaway_screen}.

Within this small sample, WR\,61, WR\,121--16, WR\,92, and WR\,119 satisfy both the tangential-velocity and height criteria, while WR\,7 is mainly supported by the velocity criterion.
WR\,58 and WR\,82 are supported primarily by the height criterion, with WR\,82 remaining close to the adopted velocity threshold.
WR\,38 does not satisfy either runaway criterion in this screen.
WR\,120 is not classified as a secure runaway because its RUWE value exceeds the adopted astrometric-quality threshold.

This result does not by itself identify the evolutionary channel of WR\,92 or WR\,119.
Runaway-like kinematics can arise from several mechanisms, including dynamical ejection, binary-supernova ejection, or more complex cluster and post-interaction histories \citep{Arthur2025}.
Nevertheless, the fact that both updated low-luminosity WC stars are kinematically unusual reinforces the view that they should not be interpreted as ordinary single-star disk objects without caution.
This is consistent with previous work showing that low-luminosity late-type WC stars are difficult to reproduce with standard single-star tracks and that close-binary evolution provides a promising channel for producing such objects \citep{Peng2022LowLWC}.
The kinematic screen therefore provides additional, although not decisive, support for considering non-standard evolutionary histories in the faint WC regime.

\section{Conclusions}
\label{sec:conclusions}

We have performed a systematic comparison between atmosphere-derived parameters of Galactic WR stars and rotating single-star evolutionary models at solar metallicity ($Z=0.02$), with particular emphasis on the low-luminosity WR population.
The comparison was designed to proceed from HR diagram proximity, to subtype/composition consistency, and finally to wind-scaling diagnostics, allowing us to distinguish between cases that remain broadly compatible with single-star evolution and those that require either additional physics or an alternative evolutionary channel.

The adoption of the enhanced RSG mass-loss prescription (S99) significantly lowers the minimum initial mass required for WR-star formation, reducing it from $25\,M_\odot$ to approximately $18\,M_\odot$. 
This not only offers a plausible formation channel for low-luminosity WNC/WC stars, but also naturally explains the presence of low-luminosity WN stars in the Galactic WR catalogue.
The enhanced mass-loss rates also help to explain the formation of H-poor and H-free supernova progenitors.

The evolutionary models predict WR-star masses in the range of $4.5$--$12\,M_\odot$. Moreover, the stellar masses derived from the best-fitting models are systematically lower than the current stellar masses inferred from a mass--luminosity (M--L) relation for homogeneous helium stars.
The ages of the low-luminosity WR stars are estimated to range from approximately \(5.5\)--\(11\,{\rm Myr}\). Most of the low-luminosity WR stars are found in the late stages of core He burning (\(Y_{\rm c}<0.4\)). This is particularly true for the WNC and WC stars, which are generally located very close to the exhaustion of central helium.

Although the S20 prescription is calibrated using observed WR-star mass-loss rates and stellar atmosphere models through its dependence on $\Gamma_{\rm e}$, it predicts mass-loss rates that are significantly lower than the observed values in the low-luminosity regime, especially for stars with $\log L/L_\odot \leq 5.2$.
The higher mass-loss rates predicted by the HS19 prescription during the WNC and WC phases provide a viable single-star channel for the formation of low-luminosity WC stars. These models evolve into low-mass, nearly naked CO core stars at the pre-supernova stage, making them potential low-luminosity progenitor candidates of Type Ic supernovae. Although our evolutionary grids can reproduce the luminosities and overall distributions of most low-luminosity WR stars, the predicted WR models remain systematically too hot. Consequently, they cannot fully explain the observed properties of the cooler WNE, WNC, and WCL samples, indicating that the long-standing ``temperature problem'' of WR stars remains unresolved.

 
Among the nine objects, WR\,92 and WR\,119 satisfy both the tangential-velocity and height runaway criteria in the Gaia-based kinematic screen, marking them as kinematically unusual disk objects even without invoking a specific formation channel.
Together with the model--data comparison, this supports treating the lowest-luminosity WC stars as potentially non-standard evolutionary products, consistent with previous work suggesting that close-binary evolution is a promising route to low-luminosity late-type WC stars.

Overall, low-luminosity WNC and WC stars provide especially sensitive tests of WR mixing, WR-phase mass loss, and the contribution of binary products to the Galactic WR population. Progress on this problem will require both improved control of observational systematics and tighter constraints on the range of post-interaction helium-star channels that can produce genuinely WR-like spectra.

\begin{acknowledgements}
This work is supported by the China Scholarship Council (CSC).
This work is also supported by the National Key R\&D Program of China (grant No. 2021YFA1600400/2021YFA1600402), the National Natural Science Foundation of China (Nos. 12133011 and 12288102), and the International Centre of Supernovae, Yunnan Key Laboratory (No. 202302AN360001). 
This work is supported by the China Manned Space Program with grant no. CMS-CSST-2025-A14.
AJCT acknowledges the Spanish Ministry of Science, Innovation and Universities project PID2023-151905OB-I00. MG thanks Adolfo González Rivera (Alhama Academy) for logistical support and acknowledges funding from the Academy of Finland (project no. 325806). The programme of development within Priority-2030 is acknowledged for supporting the research at UrFU. MCG acknowledges financial support from the Spanish Ministry project MCI/AEI/PID2023-149817OB-C31.
\end{acknowledgements}


\bibliographystyle{aa}
\bibliography{reference}




\end{document}